\documentclass{aa}
\usepackage{graphicx}
\usepackage{txfonts}
\usepackage{mhchem}
\usepackage{xcolor}
\usepackage{booktabs}
\usepackage{siunitx}
\usepackage{utfsym} % to add crossmarks
\usepackage{hyperref}
\begin{document}

    \title{Conformational isomerism of methyl formate: new detections of the higher-energy {\it trans} conformer and theoretical insights}
    \authorrunning{Sanz-Novo et al.}\titlerunning{Conformational isomerism of methyl formate}
   \author{M. Sanz-Novo
          \inst{1}, G. Molpeceres
          \inst{2}, V. M. Rivilla \inst{1},
          I. Jimenez-Serra \inst{1}
          }
    \institute{Centro de Astrobiolog{\'i}a (CAB), INTA-CSIC, Carretera de Ajalvir km 4, Torrej{\'o}n de Ardoz, 28850 Madrid, Spain
        \and
         Departamento de Astrof{\'i}sica Molecular, Instituto de F{\'i}sica Fundamental (IFF-CSIC), Madrid 28006, Spain.
    }

\date{\today}
 
  \abstract
  % context heading
  {In recent astrochemical studies it has become essential to study not only the most stable conformers but also all the structures within the conformational panorama of the molecule, some of which are potentially detectable in the interstellar medium (ISM). In this context, the isomeric ratio can be used as a powerful tool to distinguish between different formation routes of molecules with increasing levels of complexity. }
  % aims heading (mandatory)
   {While the most stable \textit{cis}-conformer of methyl formate (\ce{CH3OCHO}) is ubiquitous in the ISM, there is just one tentative detection of the higher-energy \textit{trans} form ($\Delta$$E$ = 3000 K) toward the envelope of the star-forming region Sgr B2(N). In this work, we aim to search for \textit{trans}-methyl formate in additional sources as well as to perform new theoretical computations to better understand its conformational isomerism.}
  % methods heading (mandatory)
   {We use an ultradeep molecular line survey of the Galactic  Center molecular cloud G+0.693-0.027, carried out with the Yebes 40m and IRAM 30m telescopes, as well as publicly available data from the Large Program ASAI observed toward the prototypical protostellar shock L1157-B1. The observational results are compared with predictions based on new grain-surface theoretical computations, which are sensitive to the stereochemistry of the molecule.}
  % results heading (mandatory)
   {We present the detections of \textit{trans}-methyl formate in both astronomical regions, providing conclusive observational evidence of its presence in the ISM.  Numerous unblended or slightly blended $a$-type $K_a$ = 0, 1 transitions belonging to the $A$-substate of \textit{trans}-methyl formate have been identified in both sources, many of which have been directly observed in radio astronomical data for the first time and remain unmeasured in the laboratory. We derive a molecular column density for \textit{trans}-methyl formate of $N$ = (8.2 $\pm$ 0.4) $\times$10$^{12}$ cm$^{-2}$ and $N$ = (1.6 $\pm$ 0.3) $\times$10$^{12}$ cm$^{-2}$, respectively, for each source, yielding a molecular abundance with respect to H$_2$ of $\sim$6 $\times$ 10$^{-11}$ and $\sim$8 $\times$ 10$^{-10}$, respectively. Therefore, we obtain \textit{cis}/\textit{trans} isomeric ratios of $\sim$72 and $\sim$34 toward G+0.693 and L1157-B1, respectively, which are $\sim$7 and 3 times larger than that found in the Sgr B2(N) region. The results from our theoretical computations suggest that a stereoespecific formation of \textit{trans}-methyl formate via the \ce{CH3O + HCO} route on grain surfaces can qualitatively explain the observed \textit{cis}/\textit{trans} abundance ratio. Nevertheless, we show that additional stereoespecific gas-phase routes could also play a crucial role in maintaining the intricate balance between formation and destruction of \textit{trans}-MF, ultimately leading to its detection.   
   } 
   % conclusions heading (optional), leave it empty if necessary 
   {}

    \maketitle

\section{Introduction}
\label{intro}

The study of isomerism in the interstellar medium (ISM) stands as a prolific field in recent astrochemical studies. However, the detection of high-energy isomers in the ISM has historically faced skepticism, due to their expected low abundances compared to their more stable counterparts under the interstellar conditions \citep{lattelais2009}. Recent findings challenge this principle for various molecular systems, highlighting several families of structural isomers\footnote{Molecules with the same molecular formula but a different bonding arrangement among the atoms.} (e.g., the \ce{C3H2O}, \ce{C2H4O2}, \ce{C2H2N2} or \ce{C2H5O2N} isomeric families; \citealt{shingledecker2019,mininni2020,Rivilla2023,SanAndres2024}).

On the contrary, several stereoisomers\footnote{Molecules with the same molecular formula and connectivity of atoms, but different 3D-arrangement in space.}, such as both $Z$- and $E$- stereoisomers of C-cyanomethanimine (HNCHCN; \citealt{SanAndres2024}), the \textit{trans} and \textit{gauche} conformers\footnote{Simplest example of stereoisomerism, molecules differing only in the rotations about single bonds.} of ethanol (\ce{C2H5OH}; \citealt{Pearson:1997ki}), the \textit{Ga} and \textit{Aa} conformers of \textit{n}-propanol (\textit{n}-\ce{C3H7OH}; \citealt{jimenez-serra2022}), the \textit{anti} and \textit{gauche} conformers of ethyl formate (\ce{CH3CH2(O)CHO}; \citealt{rivilla_chemical_2017}), and also the \textit{cis} and \textit{trans} conformers of thioformic acid (HC(O)SH; \citealt{garciadelaconcepcion2022}) support the idea that relative abundances correlate with molecular stability. Nonetheless, even some conformers, including \textit{cis-cis} and \textit{cis-trans} carbonic acid (HOCOOH; \citealt{Sanz-Novo2023}), the \textit{cis} and \textit{trans} conformers of formic acid (HCOOH; \citealt{garciadelaconcepcion2022}), and methyl formate (\ce{HC(O)OCH3}; \citealt{Neill:2012fr,Faure:2014iu}), show that this correlation is not absolute and that thermodynamic factors alone can not rationalize the observed abundance ratios \citep{Molpeceres2022}.

\begin{center}
\begin{figure}[ht]
     \centerline{\resizebox{0.8
     \hsize}{!}{\includegraphics[angle=0]{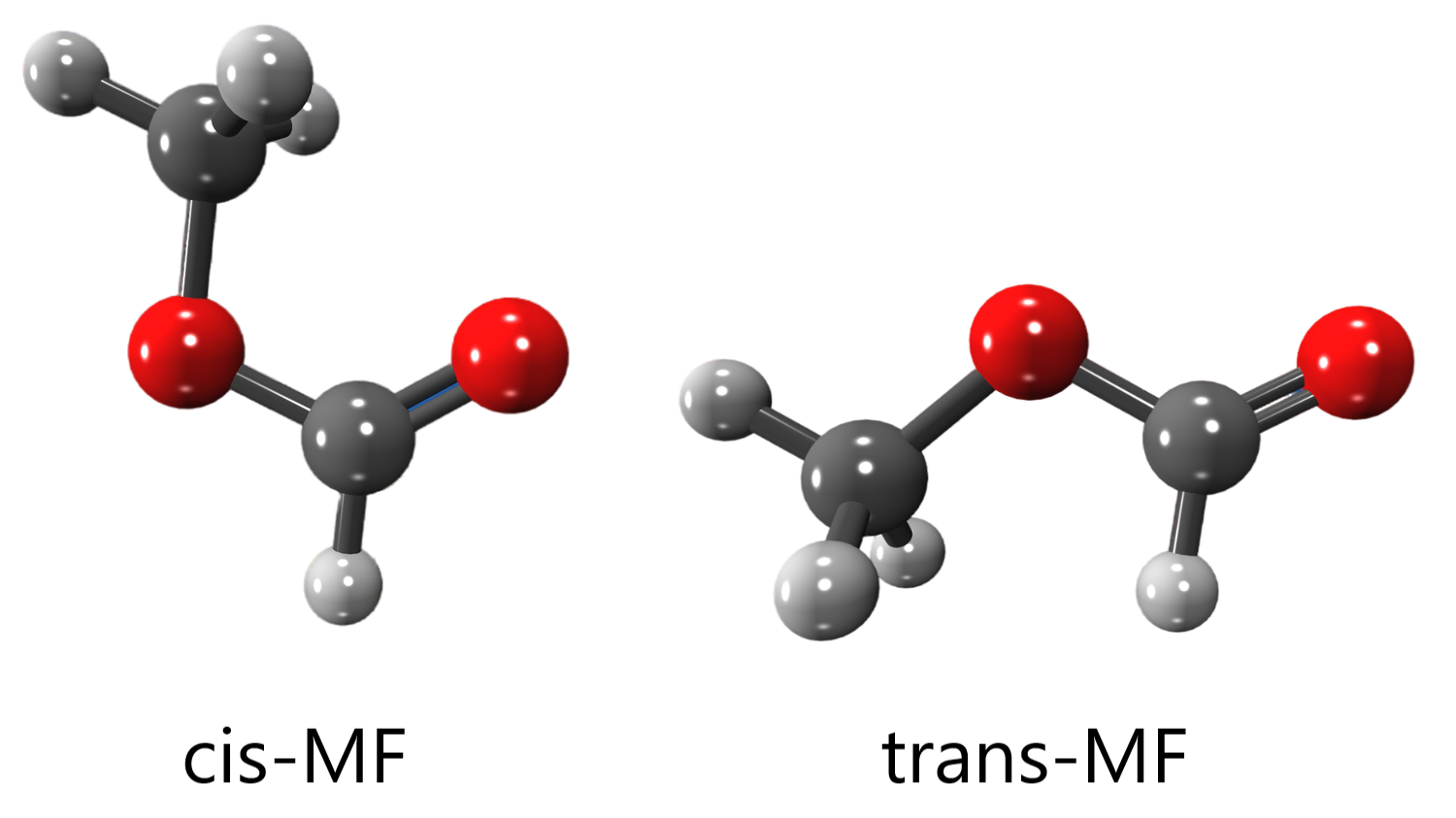}}}
     \caption{Structures of the lowest-energy conformers of methyl formate (MF), separated by a $\Delta$$E$ = 3000 K (25 kJ mol$^{-1}$). Color code: carbon atoms in gray, oxygen atoms in red, and hydrogen atoms in white.}
\label{f:3D}
\end{figure}
\end{center}

Methyl formate (hereafter MF), a molecule that is ubiquitous in the ISM, appears as a promising testbed to study stereoisomerism in space. The \textit{cis} conformer (see Fig. \ref{f:3D}), often referred to solely as MF, was firstly detected in emission toward Sgr B2 \citep{Brown:1975jh}. Afterward, this species has been detected in a plethora of interstellar environments, including prestellar cores (e.g., \citealt{bacmann_detection_2012,Jimenez-Serra16,Megias2023,Scibelli2024}), hot cores (both in the inner and outer galaxy; see e.g., \citealt{belloche_complex_2013,rivilla_chemical_2017,mininni2020,Shimonishi2021}), hot corinos \citep{cazaux_hot_2003,pineda2012}, molecular clouds (e.g., \citealt{requena-torres_organic_2006,Agundez2021}), in protoplanetary disks \citep{Lee2019,Brunken2022}) and even in extragalactic objects (e.g., \citealt{Sewiio:2018jx}).

Meanwhile, the so-called \textit{trans} conformer, with the carbonyl oxygen pointing away from the methyl group (see Fig. \ref{f:3D}), is much higher in energy than the \textit{cis} conformer (about 25 kJ mol$^{-1}$ or $\sim$3000 K), and to date, it has only been detected tentatively toward the Sagittarius B2(N) molecular cloud, based on the identification of absorption features \citep{Neill:2012fr}.

In this context, and motivated by the recent detection of several high-energy stereoisomers towards the Galactic Center (GC) molecular cloud G+0.693-0.027 (hereafter G+0.693, see e.g., \citet{jimenez-serra2022,Sanz-Novo2023,Rivilla2023,SanAndres2024}), we have searched for \textit{trans}-MF, yielding a clear and positive detection in emission. This detection toward this source, which is dominated by large-scale shocks \citep{requena-torres_organic_2006,zeng2020} that sputter the dust grains and increase the gas-phase molecular abundances, encouraged us to search for this species toward other shocked regions. In particular, we have chosen the archetypal protostellar shock L1157-B1, which shows a rich chemistry (e.g., \citealt{Gusdorf:2008ez,Codella:2015kq,lefloch_chess_2012,busquet2014,Holdship:2016js,lefloch_l1157-b1_2017}), including also the presence of \textit{trans}-MF.

While substantial research has been conducted on the interstellar formation of MF, considering both gas-phase and grain-surface mechanisms (e.g., \citealt{Horn2004,Neill:2011pd,laas2011,Cole2012,Balucani:2015gj} and \citealt{Garrod:2006gw,garrod2013ApJ,enrique-romero_quantum_2022}), most studies -particularly those focused on grain-surface processes- predominantly focus on the formation of the most stable \textit{cis} conformer. This has resulted in lack of understanding regarding the formation of \textit{trans}-MF and its expected isomeric ratio. Therefore, to fill this gap, the observational results will also be compared with predictions based on new grain-surface theoretical computations, which are sensitive to the stereochemistry of the molecule.

\section{Interstellar searches}
\label{sec:obs}

\subsection{Observations} 
\label{subsec:obs}

We used an unbiased and ultradeep spectral survey performed towards the GC molecular cloud G+0.693 to search for \textit{trans}-MF. This source is an  intermediate-dense cloud (a few 10$^4$ cm$^{-3}$) with warm kinetic temperatures (about 70-150 K), as described in \citet{zeng2018} and \citet{Colzi2024}. Also, it is characterized by exhibiting low dust temperatures ($\sim$20--30 K in the Galactic Center molecular clouds; \citealt{rodriguez-fernandez2000,Etxaluze2013}). Moreover, G+0.693 appears as an astrochemical mine for the discovery of new species in the ISM, with more than 20 first detection over the last few years (see e.g., \citealt{rivilla2019b,rivilla2020b,rivilla2021a,rivilla2021b,rivilla2022a,rivilla2022b,Rivilla2023,rodriguez-almeida2021a,rodriguez-almeida2021b,jimenez-serra2022,zeng2021,zeng2023,SanAndres2024,Sanz-Novo2023,Sanz-Novo2024a,Sanz-Novo2024b,Sanz-Novo2025}).

We conducted observations in the $Q$-band (31.075-50.424 GHz) using the Yebes 40$\,$m radiotelescope located in Guadalajara (Spain). Additionally, we covered three more frequency windows with high sensitivity using the IRAM 30$\,$m radiotelescope in Granada (Spain): 83.2$-$115.41 GHz, 132.28$-$140.39 GHz, and 142.00$-$173.81 GHz. We used the position switching mode, centered at $\alpha$ = $\,$17$^{\rm h}$47$^{\rm m}$22$^{\rm s}$, $\delta$ = $,-$28$^{\circ}$21$^{\prime}$27$^{\prime\prime}$, with the off position shifted by $\Delta\alpha$=$-885^{\prime\prime}$ and $\Delta\delta$~=~$290^{\prime\prime}$. The half power beam width (HPBW) of the Yebes 40$\,$m telescope ranges from approximately 35$^{\prime\prime}$ to 55$^{\prime\prime}$ at frequencies of 50 GHz and 31 GHz, respectively \citep{tercero2021}. In contrast, the HPBW of the IRAM 30$\,$m radiotelescope varies from $\sim$14$^{\prime\prime}$ to $\sim$29$^{\prime\prime}$ across the covered frequency range. Further details on these observations, such as resolution and noise levels of the molecular line survey, are provided in \citet{Rivilla2023} and \citet{Sanz-Novo2023}.

We also searched for \textit{trans}-MF toward L1157-B1, a bright bow-shock in the powerful and chemically rich outflow powered by the L1157 low-mass protostar \citep{Bachiller1997,Bachiller:2001ju}, using the publicly available data from the Large Program ASAI (Astrochemical Surveys At IRAM; \citealt{Lefloch18}). These data are based on IRAM 30-m observations, using the broad-band EMIR receivers connected to Fast Fourier Transform Spectrometers in the 195 kHz spectral resolution mode. In particular, we explored the data covering two spectral windows: 71.7-79.5 GHz and 78.8-118.0 GHz, the latter with a frequency resolution of 390 kHz (which translate to velocities of 1.5 and 1 km s$^-1$, respectively). The source nominal positions are $\alpha$$_{J2000}$ = 20$^{\rm h}$ 39$^{\rm m}$ 10$^{\rm s}$, $\delta$$_{J2000}$ = +68$^{\circ}$ 01$^{\prime}$ 10$^{\prime\prime}$. The cloud ambient velocity $v$$_{\rm LSR}$ = +2.6 km s$^{-1}$. We refer to \citealt{Lefloch18} for a complete description of the observations and data acquisition. Fluxes are expressed in main beam temperature units ($T$$_{mb}$), as the source does not fill the beam. The beam efficiency, $B$$_{eff}$ and forward efficiency, $F$$_{eff}$) values were taken from the IRAM webpage\footnote{IRAM; \url{https://publicwiki.iram.es/Iram30mEfficiencies}}.

\begin{center}
\begin{figure*}[ht]
     \centerline{\resizebox{0.7
     \hsize}{!}{\includegraphics[angle=0]{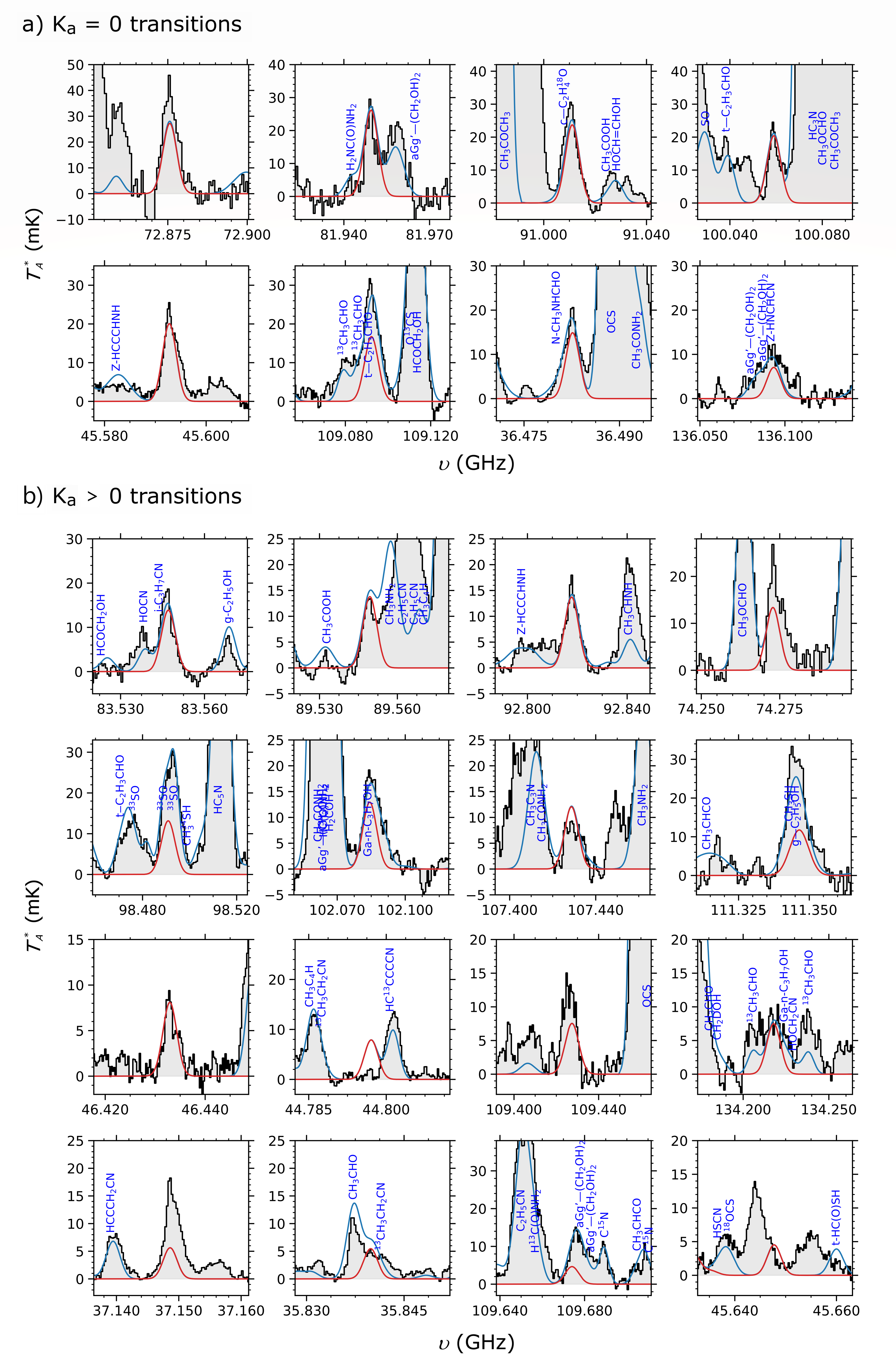}}}
     \caption{Transitions of \textit{trans}-MF ($A$-symmetry substate) identified toward G+0.693 (listed in Table \ref{tab:transitions}) sorted by decreasing peak intensity. The result of the best LTE fit of \textit{trans}-MF is plotted with a red line and the blue line depicts the emission from all the molecules identified to date in our survey of G+0.693 including \textit{trans}-MF (observed spectra shown as gray histograms).}
\label{f:LTEspectrumtransG0693}
\end{figure*}
\end{center}

\subsection{Conformational panorama of MF and rotational spectroscopy considerations}
\label{subsec:rot}

The conformational panorama of MF exhibits two plausible configurations, the \textit{cis} conformer (energy minimum), and the \textit{trans} conformer (located at 24.6 kJ mol$^{-1}$, or 2980 K higher in energy; \citealt{Neill:2012fr}, see Figure \ref{f:3D}). These results were obtained experimentally based on relative intensity measurements of the 1$_{0,1}$ -- 0$_{0,0}$ transitions belonging to the $A$-symmetry substate of each conformer \citep{Neill:2012fr}. Additionally, both conformers are asymmetric tops exhibiting methyl internal rotation motions due to the presence of a -CH$_3$ group. This motion is less hindered for \textit{trans}-MF, exhibiting a three-fold ($V$$_3$) potential barrier (12.955 $\pm$ 0.017 cm$^{-1}$; \citealt{Neill:2012fr}) that is considerably lower than that obtained for \textit{cis}-MF  (379.4 cm$^{-1}$; \citealt{Karakawa2001}). This lower barrier hampers its rotational characterization due to a larger expected A-E splitting. Nevertheless, \textit{trans}-MF shows a much larger $\mu$$_a$ dipole moment component (4.2 D; \citealt{Neill:2012fr}) compared to \textit{cis}-MF (1.6 D; \citealt{Curl1959}), a crucial factor that enables its detection both in the laboratory and in space.

\begin{table*}
\centering
\tabcolsep 4.5pt
\caption{List of observed transitions of \textit{trans}-MF ($A$-symmetry substate) detected toward G+0.693 and L1157-B1 that are unblended or slightly blended. We provide the transitions frequencies, quantum numbers, base 10 logarithm of the integrated intensity at 300 K (log $I$), and the values of the upper levels ($E_{\rm u}$) and degeneracy of each transition.}
\begin{tabular}{ c c c c c c c l}
\hline
 Frequency & Transition  & log $I$ & $E_{\rm u}$  & $g_{\rm u}$ & G+0.693 & L1157-B1 & Blending (G+0.693/L1157-B1)$^a$ \\
 (GHz) & ($J_{K_a,K_c}$)  &   (nm$^2$ MHz)  & (K) &   \\
\hline
35.8400774 & 4$_{1,4}$ - 3$_{1,3}$ & -4.6278   &  7.1 & 9 & $\checkmark$ &  \scalebox{0.85}{\usym{2613}} & $^{13}$\ce{CH3CH2CN} and \ce{CH3CHO} /  \scalebox{0.85}{\usym{2613}} \\ %xx  
36.4828481 & 4$_{0,4}$ - 3$_{0,3}$ & -4.5803   &  4.3 & 9 & $\checkmark$ &  \scalebox{0.85}{\usym{2613}} & \ce{N-CH3NHCHO} /  \scalebox{0.85}{\usym{2613}} \\   %xx
37.1488320 & 4$_{1,3}$ - 3$_{1,2}$ & -4.5968   &  7.2 & 9 & $\checkmark$ &  \scalebox{0.85}{\usym{2613}} &  U-line /  \scalebox{0.85}{\usym{2613}} \\ %xx
44.7973533 & 5$_{1,5}$ - 4$_{1,4}$ & -4.3296   &  9.2 & 11 & $\checkmark$ &  \scalebox{0.85}{\usym{2613}} & Unblended /  \scalebox{0.85}{\usym{2613}} \\  %xx
45.5930371 & 5$_{0,5}$ - 4$_{0,4}$ & -4.2927   &  6.5 & 11 & $\checkmark$ &  \scalebox{0.85}{\usym{2613}} & Unblended /  \scalebox{0.85}{\usym{2613}} \\    %xx
45.6480015 & 5$_{2,3}$ - 4$_{2,2}$ & -4.3835   & 17.6 & 11 & $\checkmark$ &  \scalebox{0.85}{\usym{2613}} & U-line /  \scalebox{0.85}{\usym{2613}}\\ %xx
46.4332081 & 5$_{1,4}$ - 4$_{1,3}$ & -4.2987   & 16.0 & 11 & $\checkmark$ &  \scalebox{0.85}{\usym{2613}} &  Unblended /  \scalebox{0.85}{\usym{2613}}\\ %xx  
72.8760742* & 8$_{0,8}$ - 7$_{0,7}$ & -3.6936   & 12.2 & 17 & $\checkmark$ &  \scalebox{0.85}{\usym{2613}} & U-line /  \scalebox{0.85}{\usym{2613}}\\  %xx 
74.2732762* & 8$_{1,7}$ - 7$_{1,6}$ & -3.6883   & 18.7 & 17 & $\checkmark$ &  \scalebox{0.85}{\usym{2613}} & U-line /  \scalebox{0.85}{\usym{2613}}\\  %xx 
80.6050090* & 9$_{1,9}$ - 8$_{1,8}$  & -3.5692  &  22.0 & 19 & $\checkmark$ & $\checkmark$ & Heavily blended / Unblended \\  %xx  \\ 
81.9500446* & 9$_{0,9}$ - 8$_{0,8}$ & -3.5459  &  19.5 & 19 & $\checkmark$ & $\checkmark$ & Unblended / Unblended\\    %xx
83.5475707* & 9$_{1,8}$ - 8$_{1,7}$ & -3.5390   & 22.7 & 19 & $\checkmark$ &  \scalebox{0.85}{\usym{2613}} &  i-\ce{C3H7CN} /  \scalebox{0.85}{\usym{2613}}  \\ %xx  
89.5499417* & 10$_{1,10}$ - 9$_{1,9}$ & -3.4369  &  26.3 & 21 & $\checkmark$ & $\checkmark$ & CH$_3$NH$_2$ / Unblended\\  %xx  \\  \\  
91.0116435* & 10$_{0,10}$ - 9$_{0,9}$ & -3.4150  &  23.9 & 21 & $\checkmark$ & $\checkmark$ &  \textit{c}-C$_2$H$_4$$^{18}$O / Unblended  \\ %xx  
92.8182813* & 10$_{1,9}$ - 9$_{1,8}$ & -3.4069   & 27.1 & 21 & $\checkmark$ & \scalebox{0.85}{\usym{2613}} &  Unblended /  \scalebox{0.85}{\usym{2613}} \\   %xx 
98.4914827* & 11$_{1,11}$ - 10$_{1,10}$ & -3.3186   & 31.0 & 23 & $\checkmark$ &  \scalebox{0.85}{\usym{2613}} & $^{33}$SO, CH$_3$$^{34}$SH /  \scalebox{0.85}{\usym{2613}}\\  %xx 
100.0596405* & 11$_{0,11}$ - 10$_{0,10}$ & -3.2980  &  28.7 & 23 & $\checkmark$ & $\checkmark$ & Unblended / Unblended \\ %xx  
102.0849526* & 11$_{1,10}$ - 10$_{1,9}$ & -3.2889  &  32.0 & 23 & $\checkmark$ & $\checkmark$ &  \textit{Ga-n}-\ce{C3H7OH} / Unblended \\  %xx
107.4293606* & 12$_{1,12}$ - 11$_{1,11}$ & -3.2120  &  36.1 & 25 & $\checkmark$ & $\checkmark$ &  \textit{Ga-n}-\ce{C3H7OH} / Unblended\\  %xx
109.0928847* & 12$_{0,12}$ - 11$_{0,11}$ & -3.1924  &  33.9 & 25 & $\checkmark$ & $\checkmark$ &  \textit{t}-\ce{C2H3CHO} / Unblended \\ %xx 
109.4280533* & 12$_{2,11}$ - 11$_{2,10}$ &  -3.2181  &  45.0 & 25 & $\checkmark$ &  \scalebox{0.85}{\usym{2613}} &  U-line /  \scalebox{0.85}{\usym{2613}}\\   
109.6747106* & 6$_{1,6}$ - 5$_{0,5}$ &  -4.1351  &   11.7 & 13 & $\checkmark$ & \scalebox{0.85}{\usym{2613}} &   \textit{aGg'}-\ce{(CH2OH)2} / \scalebox{0.85}{\usym{2613}}\\  %xx
111.3471110* & 12$_{1,11}$ - 11$_{1,10}$ & -3.1825  &  37.3 & 25 & $\checkmark$ &  $\checkmark$ &   \textit{g}-\ce{C2H5OH} / Unblended \\  %xx  
134.2185890* & 15$_{1,15}$ - 14$_{1,14}$ &  -2.9458  &  54.0 & 31 & $\checkmark$ &  \scalebox{0.85}{\usym{2613}} &  \textit{Ga-n}-\ce{C3H7OH} and $^{13}$\ce{CH3CHO} /  \scalebox{0.85}{\usym{2613}} \\ %xx  
136.0942706* & 15$_{0,15}$ - 14$_{0,14}$ &  -2.9290  &  52.0 & 31 & $\checkmark$ &  \scalebox{0.85}{\usym{2613}} &  \textit{Z}-HNCHCN \\  %xx
\hline 
\end{tabular}
\label{tab:transitions}
{\\ (a) ``U" refers to blendings with emission from an unknown (not identified) species. Lines that have been observed for the first time in the present astronomical datasets are marked with a * symbol. Transitions that have not been measured toward L1157-B1 -either because the frequency range was not covered (e.g., $Q$-band data) or the features do not arise within the noise- are marked with a \scalebox{0.85}{\usym{2613}} symbol.
}
\end{table*}

\begin{table*}
\centering
\caption{Derived physical parameters for \textit{cis}- and \textit{trans}-MF toward the G+0.693 molecular cloud and L1157-B1.}
\begin{tabular}{ c c c c c c c  }
\hline
\hline
 Source & Molecule  & $N$   &  $T_{\rm ex}$ & $v$$_{\rm LSR}$ & FWHM  & Abundance$^a$   \\
 & & ($\times$10$^{13}$ cm$^{-2}$) & (K) & (km s$^{-1}$) & (km s$^{-1}$) & ($\times$10$^{-10}$)  \\
\hline
G+0.693   &  \textit{cis}-MF & 60 $\pm$ 3 & 13.4 $\pm$ 0.8 & 68.3 $\pm$ 0.8 & 21$^b$  & 44 $\pm$ 7     \\
          &  \textit{trans}-MF ($K$$_a$ = 0) & 0.82 $\pm$ 0.04  & 14 $\pm$ 4 & 70.6 $\pm$ 2.5 & 21$^b$ & 0.61 $\pm$ 0.09 \\
\hline
L1157-B1$^c$ & \textit{cis}-MF  &  5.5 $\pm$ 0.2 & 15.7 $\pm$ 0.6 & -0.3 $\pm$ 0.1 & 5.5 $\pm$ 0.2 & 280 $\pm$ 40  \\
     & \textit{trans}-MF & 0.16 $\pm$ 0.02 & 14 $\pm$ 3 & -0.6 $\pm$ 0.3 & 5.2 $\pm$ 0.6 &  8 $\pm$ 2  \\
\hline 
\end{tabular}
\label{tab:comparison}
\vspace{0mm}
{\\$^a$ We adopted $N_{\rm H_2}$ = 1.35$\times$10$^{23}$ cm$^{-2}$ for G+0.693, from \citet{martin_tracing_2008}, assuming an uncertainty of 15\% of its value, while for L1157-B1, we adopted a  $N_{\rm H_2}$ = 2$\times$10$^{21}$ cm$^{-2}$, from \citet{lefloch_chess_2012,DiFrancesco2020}, assuming an uncertainty of 15\% of its value. $^b$ Value fixed in the fit. $^c$ For both MF conformers we assumed a source size of 18'' as that derived for the strong shock tracer SiO \citep{Lefloch2016}.}
\label{tab:g0693}
\end{table*}

Moreover, while extensive spectroscopic data is available in the literature for \textit{cis}-MF (\citealt{Ilyushin:2009kg}, and references therein), the rotational spectroscopy of \textit{trans}-MF is limited to the 6-60 GHz frequency range \citep{Neill:2012fr}. To date, 24 transitions have been measured using several pulsed-jet FTMW spectrometers for the $A$-symmetry substate, while only 9 (up to 36 GHz) belonging to the $E$-substate \citep{Neill:2012fr}. Consequently, we have used exclusively the available transition frequencies of the $A$-symmetry lines to perform an effective least-squares fit to a semi-rigid rotor Hamiltonian (Watson's $A$-reduced Hamiltonian in $I$$^r$ representation). The resulting spectroscopic constants are listed in Table \ref{t:mwtable1}. We then extrapolated these results to higher frequencies to include the spectral coverage of the observations we have used here, where the brightest transitions of \textit{trans}-MF are expected (according to the low $T_{\rm ex}$ in both sources), and prepared the corresponding line catalogs. Note that even though we anticipate non-negligible uncertainties upon reaching the millimeter-wave region, these uncertainties would not have an impact on the analysis, specially for transitions with low values of $K$$_a$, as they will likely be notably smaller than the typical line widths of the molecular line emission observed toward G+0.693 but also L1157-B1 (e.g., $<$100 kHz or 0.2 km s$^{-1}$ for the 12$_{1,12}$ - 12$_{1,11}$ transition at 111.347 GHz, and  $<$150 kHz or 0.33 km s$^{-1}$ for the 15$_{0,15}$ - 14$_{0,14}$ transition at 136.094 GHz compared to a FWHM of $\sim$15$-$20 km s$^{-1}$ and $\sim$5 km s$^{-1}$ for G+0.693 and L1157-B1, respectively; \citealt{requena-torres_organic_2006,requena-torres_largest_2008,zeng2018,lefloch_l1157-b1_2017,Lefloch18}). Indeed, a similar extrapolation approach, though initially applied to a much more limited experimental dataset (i.e., only three $K$$_a$ = 0 transitions), was used for the detection of HOCS$^+$ \citep{Sanz-Novo2024a}, and has recently been demonstrated to work properly for cold sources \citep{Lattanzi2024}. In particular, we found that the uncertainties in the rest frequencies were a factor of 2 lower than those expected (e.g., discrepancies between the extrapolated and the new frequency uncertainties measured at $\sim$103.076 GHz of 147 kHz).

\subsection{Detection of \textit{trans}-MF toward G+0.693 and derivation of the cis/trans ratio}
\label{s:observationsG0693}

\begin{center}
\begin{figure*}[ht]
     \centerline{\resizebox{0.85
     \hsize}{!}{\includegraphics[angle=0]{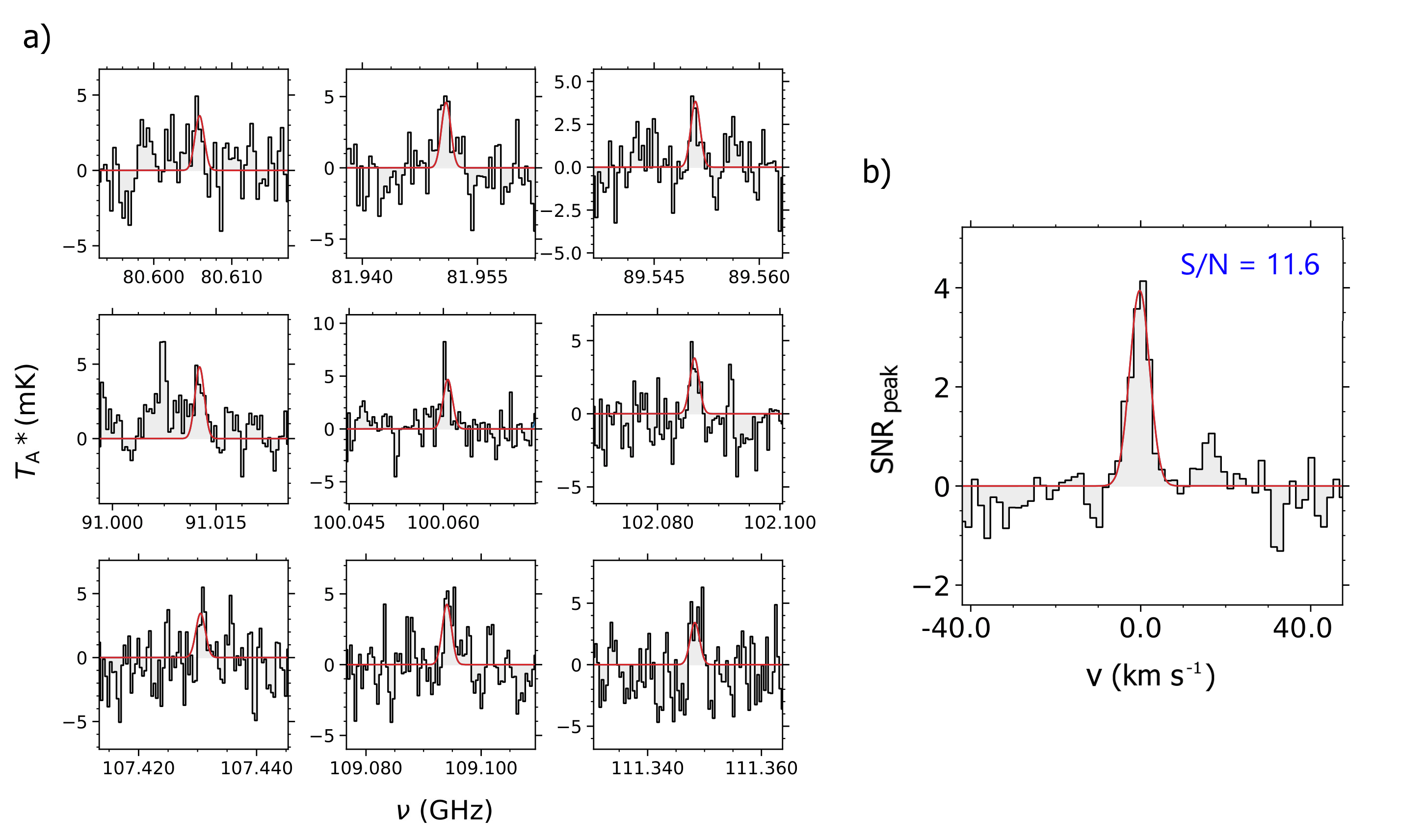}}}
     \caption{a) Transitions of \textit{trans}-MF identified ($A$-symmetry substate) toward L1157-B1 (listed in Table \ref{tab:transitions}) sorted by increasing frequency. The result of the best LTE fit of \textit{trans}-MF is plotted with a red line (observed spectra shown as black lines and gray-shaded histograms). b) Velocity stack of the observations (in gray) with the corresponding stack of the LTE simulation using the best-fit parameters to the individual lines (in red). The RMS of the stacked spectrum is 0.36 mK, implying an integrated signal-to-noise (S/N) ratio of 11.6.}
\label{f:LTEspectrumtransL1157}
\end{figure*}
\end{center}

We implemented the catalog of \textit{trans}-MF into the MADCUBA package{\footnote{Madrid Data Cube Analysis on ImageJ is a software developed at the Center of Astrobiology (CAB) in Madrid; http://cab.inta-csic.es/madcuba/; version from 2024 June 15}}\citep{martin2019} and used the Spectral Line Identification and Modeling (SLIM) tool to generate a synthetic spectra assuming Local Thermodynamic Equilibrium (LTE) excitation conditions. Although the intermediate H$_2$ volume densities of G+0.693 (of a few times 10$^4$ cm$^{-3}$; \citealt{zeng2018,Colzi2024}) result in a subthermal excitation of the molecular emission (i.e., $T_{\rm ex}$ = 5–20 K $\ll$ $T_{\rm kin}$ = 70–140 K; see, e.g., \citealt{requena-torres_organic_2006, zeng2018}), LTE models -which assume that the population of the energy levels can be described by a single temperature (commonly referred to as excitation temperature, $T_{\rm ex}$)- successfully reproduce the emission for the vast majority of species. As demonstrated by \citet{goldsmith1999}, they account for the molecular emission due to the so-called quasi-thermal excitation, a regime in which LTE is achieved despite $T_{\rm ex}$ remaining significantly lower than $T_{\rm kin}$.

In Table \ref{tab:transitions}, we list the most intense unblended or slightly blended transitions of \textit{trans}-MF ($A$-symmetry substate) observed toward G+0.693. The rest of the lines, which appear blended with brighter transitions from other species, show predicted intensities that are consistent with the observed spectra considering the contribution from more than 140 species previously identified toward G+0.693. Contrary to its tentative detection toward the envelope of Sgr B2(N), where only three $K$ = 0 transitions with $J$$_{\rm up}$ = 1-3 and a single $K$ = 1 transition belonging to the $A$-subestate were observed \citep{Neill:2012fr}, the current survey of G+0.693 allowed us to detect a myriad of $^a$$R$-branch transitions with $K$$_a$=0,1 and 2 covering from $J$$_{\rm up}$ = 4 to $J$$_{\rm up}$ = 15. Additionally, weak $^b$$R$-branch transitions (i.e., 6$_{1,5}$ -- 5$_{0,5}$) were also identified.

To derive the physical parameters of the \textit{trans}-MF emission, we used the \textsc{Autofit} tool within \textsc{MADCUBA-Slim} \citep{martin2019}, which carries out a nonlinear least-squares fitting of the simulated LTE spectra to the observed astronomical data, considering the expected emission from the already identified molecules in the same frequency region. The best-fitting LTE model for \textit{trans}-MF was achieved after splitting the transitions into the $K$$_a$ = 0 and $K$$_a$ $>$ 0 rotational ladders, as has been previously done for the analysis of other molecules toward this cloud \citep{zeng2018,rodriguez-almeida2021a}. The fitted lines profiles of \textit{trans}-MF are shown in Figure \ref{f:LTEspectrumtransG0693} (red solid line). As shown in Figure \ref{f:LTEspectrumtransG0693}, the LTE model works fine for all transitions with the only exception of the 5$_{1,5}$ - 4$_{1,4}$ transition at $\sim$44.797 GHz, which appears to be overestimated. This fact is likely due to plausible non-LTE effects, which are currently out of the scope of this work as collisional rate coefficients are not available for this high-energy conformer, leaving LTE analysis as the only viable method to determine its excitation conditions in G+0.693. Given that the number of detected clean lines is very high (with over a dozen unblended or only negligibly blended lines that perfectly fit the observed spectra; see Figure \ref{f:LTEspectrumtransG0693}), we are confident about the robust detection of this molecular species. We derived a molecular column density for the $K$$_a$=0 lines of $N$ = (8.2 $\pm$ 0.4) $\times$10$^{12}$ cm$^{-2}$, a $T_{\rm ex}$ = 14 $\pm$ 4 K and a radial velocity of $v$$_{\rm LSR}$ = 70.6 $\pm$ 2.5~km~s$^{-1}$, while the full width half maximum (FWHM) was fixed in the fit to the values found for \textit{cis}-MF (FWHM = 21 km s$^{-1}$; see below). Consequently, we obtained a molecular abundance with respect to H$_2$ of $\sim$6.1 $\times$ 10$^{-11}$, assuming a $N$(H$_{2}$) = 1.35$\times$10$^{23}$ cm$^{-2}$ from \citet{martin_tracing_2008}. 

Furthermore, it is worth noticing that, apart from the lines that fall within the $Q$-band data, the remaining transitions explored in this work have been spotted for the first time in space using the spectral survey of G+0.693, and still prevail undetected in the laboratory. Hence, despite the secure detection reported here, we urge the laboratory spectroscopic community to conduct further high-resolution rotational measurements and to expand the frequency coverage for this high-energy conformer into the millimeter- and submillimeter-wave frequency domains. These data will be key for conducting reliable searches toward could sources exhibiting very narrow linewidths ($\sim$0.5 km$^{-1}$), such as the prestellar core L1544 or the cold dark cloud TMC-1 \citep{Jimenez-Serra16,Cernicharo21}, an also toward hot corinos or even hot cores (FWHM $\sim$5 km s$^{-1}$), specially when using ALMA bands 6 or 7, where higher-in-energy transitions can be efficiently populated.

Afterward, to derive the \textit{cis}/\textit{trans} abundance ratio for MF, we present an analysis of \textit{cis}-MF toward G+0.693, previously reported by \citet{requena-torres_organic_2006}. In this context, we conducted a supplementary search using the current astronomical dataset, which allowed us to significantly increase the number of observed transitions (see Table. \ref{tab:transitionscisG0693}). We used the rotational spectroscopic data set of \textit{cis}-MF reported in \citet{Ilyushin:2009kg} (entry 60003 of the JPL catalog; \citealt{Pickett:1998cp}), which include both $A$- and $E$-substates. For the subsequent LTE analysis of \textit{cis}-MF and to constrain the excitation conditions, we selected transitions that: i) are not blended with the emission from other molecules and ii) span a substantial range of rotational energy levels ($E$$_u$ ranging from 4.3 to 54.0 K; see Table \ref{tab:transitionscisG0693}). In Figure \ref{f:LTEspectrumcisG0693} we show the result of the best LTE fit derived from \textsc{Autofit}, achieved through a two-step approach as described in \cite{SanAndres2024}. First, we constrained the FWHM using exclusively the highest signal to noise unblended transitions, obtaining a value of 21 $\pm$ 2 km s$^{-1}$. Then, we performed a second fit that includes all the transitions shown in Fig. \ref{f:LTEspectrumcisG0693}, fixing the FWHM. The derived physical parameters of \textit{cis}-MF are given in Table \ref{tab:comparison}. We obtained a $N$ = (60 $\pm$ 3) $\times$10$^{13}$ cm$^{-2}$, which translates into a fractional abundance with respect to H$_2$ of $\sim$4.5 $\times$ 10$^{-9}$, a $T_{\rm ex}$ of 13.4 $\pm$ 0.8 K, which is consistent with that found for \textit{trans}-MF, and a $v$$_{\rm LSR}$ = 68.3 $\pm$ 0.8~km~s$^{-1}$. We thus obtain a \textit{cis}/\textit{trans} abundance ratio of 72 $\pm$ 16  toward G+0.693, a factor of $\sim$7 larger than that found in the envelope of Sgr B2(N) \citep{Neill:2012fr,Faure:2014iu}.

\subsection{Detection of \textit{trans}-MF toward L1157-B1 and derivation of the cis/trans ratio}

In Figure \ref{f:LTEspectrumtransL1157}(a), we show the 9 brightest transitions of \textit{trans}-MF detected toward the prototypical shock L1157-B1 that are free from contamination from other species (listed in Table \ref{tab:transitions}). In this case, we identified two $^a$$R$-branch progressions with $K$$_a$ = 0 and 1, covering from $J$$_{\rm up}$ = 9 to $J$$_{\rm up}$ = 12 belonging to the $A$-symmetry substate. We again employed the \textsc{Autofit} tool to perform the corresponding LTE fit using all the above lines. We assumed a source size of 18'' as that derived for the strong shock tracer SiO as well as PN and PO \citep{Lefloch2016}. The best-fitting model for \textit{trans}-MF yields the following physical parameters (see Table \ref{tab:comparison}): $N$ = (1.6 $\pm$ 0.3) $\times$10$^{12}$ cm$^{-2}$, $T_{\rm ex}$ = 14 $\pm$ 3 K, $v$$_{\rm LSR}$ = -0.6 $\pm$ 0.3 ~km~s$^{-1}$ and FWHM = 5.2 $\pm$ 0.6 km s$^{-1}$. We obtained a molecular abundance with respect to molecular hydrogen of $\sim$8 $\times$ 10$^{-10}$, adopting $N$(H$_{2}$) = 2$\times$10$^{21}$ cm$^{-2}$ from \citet{lefloch_chess_2012,lefloch_l1157-b1_2017,DiFrancesco2020}, which implies that \textit{trans}-MF is an order of magnitude more abundant in L1157-B1 compared to G+0.693.
 
We also conducted a complementary stacking analysis, shown in Figure \ref{f:LTEspectrumtransL1157}(b), where the so-called "stacked" spectrum was obtained by an intensity and noise-weighted average of the data at the expected frequencies of the transitions of \textit{trans}-MF. We derived a final integrated S/N ratio\footnote{The S/N ratio is calculated from the integrated signal ($\int$ $T$$\mathrm{_A^*}$d$v$) and noise level $\sigma$ = rms $\times$ $\sqrt{\delta v \times \mathrm{FWHM}}$, where $\delta$$v$ is the velocity resolution of the spectra and the FWHM is fitted from the data} of $\sim$12, which strengthen the detection even further. 

Despite the detection of the more stable \textit{cis}-MF toward this source being already reported in \citet{lefloch_l1157-b1_2017}, we repeated the analysis for comparison purposes using \textsc{MADCUBA}. A sample list of the brightest transitions of \textit{cis}-MF are shown in Figure \ref{f:LTEspectrumcisL1157} and their spectroscopic information is given in Table \ref{tab:transitionscisL1157}. The derived physical parameters of the LTE analysis using the \textsc{Autofit} tool are reported in Table \ref{tab:comparison}, along with the data of \textit{trans}-MF. Overall, our results are consistent with those reported in the previous work ($N$ = (5.4 $\pm$ 0.8) $\times$ 10$^{13}$ and $T$$_{ex}$ = 20.7 $\pm$ 1.5; \citealt{lefloch_l1157-b1_2017}), although we derived a slightly lower value for $T$$_{ex}$. This fact may be attributed to the use of a different source size compared to the previous analysis of \textit{cis}-MF toward this region, since to our knowledge this value is not explicitly mentioned in \citet{lefloch_l1157-b1_2017}.

Based on the derived molecular abundance (see Table \ref{tab:comparison}), we obtain a $\textit{cis}$/$\textit{trans}$ MF abundance ratio of 34 $\pm$ 4 toward L1157-B1, which is a factor of $\sim$2 lower than the ratio derived toward G+0.693, and about $\sim$3 times larger than that obtained toward the envelope of Sgr B2(N).

\section{Quantum chemical calculations}
The observation of \textit{trans}-MF toward these regions raises an intriguing question: a) What are the main, and potentially selective, formation pathways of \textit{trans}-MF in the ISM? Extensive research has been conducted on the interstellar production of MF, including both gas-phase (e.g., \citealt{Horn2004,Neill:2011pd,laas2011,Cole2012,Balucani:2015gj}) and grain-surface (e.g. \citealt{Garrod:2006gw,garrod2013ApJ,enrique-romero_quantum_2022}) processes. In gas-phase chemistry, MF can be formed through the oxidation of the \ce{CH3OCH2} radical, which can be formed by several reactions involving dimethyl ether (\ce{CH3OCH3}), via: O + \ce{CH3OCH2} $\rightarrow$ \ce{HC(O)OCH3} + H.

On grain surfaces, MF is thought to form through the recombination of radicals, produced by UV radiation or cosmic ray-induced photodissociation of simple molecules, through the following reaction \citep{garrod2013ApJ,enrique-romero_quantum_2022}:

\begin{equation}
\mathrm{HCO} + \mathrm{\ce{CH3O}} \rightarrow \rm \mathrm{\ce{HC(O)OCH3}}  
\label{eq:formation}
\end{equation}

%Motivation for the new computations

Nevertheless, to the best of our knowledge, these grain-surface studies are limited to the investigation of the most stable \textit{cis} conformer, and thus lack compelling data regarding the formation of \textit{trans}-MF and the expected isomeric ratio. \cite{Neill:2012fr} and \cite{Cole2012} provided hints on several gas-phase production mechanisms by which \textit{trans}-MF could be produced at a non-thermal abundance (i.e., out of thermodynamic equilibrium) in Sgr B2(N). However, no conclusions have been reached that fully rationalize and quantify the presence of \textit{trans}-MF in the ISM, especially through grain synthesis. In this regard, new grain-surface theoretical studies should provide essential reference data for constraining models of conformer-selective formation pathways. Therefore, we have conducted an accompanying theoretical study, entailing a thorough exploration of the radical-radical reaction \ref{eq:formation} on the surface of dust grains, to analyze whether it results in a stereoisomeric excess of either of the two MF conformers.

\subsection{Computational protocol}
\label{subsec:CompMeth}

The excess of a certain isomeric form can have several origins (see Section \ref{sec:discussion}). The most straightforward one is explaining the excess relying on selective formation routes. We tested this hypothesis for MF considering its formation in the surface of dust grains through the radical-radical reaction \ref{eq:formation} (\citealt{garrod2013ApJ, enrique-romero_quantum_2022}), considered to be the most important one for the molecule.

To check whether this route produces an excess of any conformer of MF, we carried sampling calculations of HCO radicals reacting with pre-adsorbed \ce{CH3O} on amorphous solid water (ASW) clusters. The reason why HCO is considered the mobile radical in the reaction is the higher binding energy (BE) of \ce{CH3O} over \ce{HCO}, making the latter prone to diffuse and react. \citet{enrique-romero_quantum_2022} showed that precaution is advised when studying this reaction on ices, due to barriers emerging as a result of the interaction of the adsorbates with ASW, and therefore we follow their exact same theoretical framework to carry out our simulations. All calculations in this work use the \textsc{Orca 5.0.4} code \citep{Neese2012, Neese2020} and our own sampling routines.

The protocol for the set-up of the calculations is as follows. We begin extracting the Cartesian coordinates of the 18 \ce{H2O} ice used in \citet{enrique-romero_quantum_2022}\footnote{\url{https://zenodo.org/records/5723996}}, and optimizing the structure under \textsc{Orca} standard convergence criteria. As mentioned above, the theoretical method used in our calculations is exactly the same as in \citet{enrique-romero_quantum_2022}, i.e. BHLYP-D3BJ/6-31+G(d,p), that was originally benchmarked against high-level multireference calculations.\footnote{We note that BHLYP-D3BJ/6-31+G(d,p) in this work and \citet{enrique-romero_quantum_2022} is slightly different because of the different standards of the codes employed. For example \textsc{Orca} uses resolution-of-identity techniques as default that depend on auxiliary basis sets. Deviations are expected in the order of $\mu$E$_{h}$, ca. 1$\times$10$^{-3}$ kcal mol$^{-1}$, in the evaluation of electronic energies.} Once the ice cluster model structure is optimized, we localized different binding sites on the cluster, placing 25 \ce{CH3O} radicals around the cluster and optimizing the initial structure with loose convergence criteria. The reason to use such convergence criteria for the optimization is to reduce the cost associated with the sampling using a sequential optimization approach (\textit{vide infra}). After this second optimization we extracted three binding configurations, sporting electronic binding energies (BE) of $\sim$ 4800 K, 2800 K and 1000 K. We call these sites High, Medium and Low attending to these BE.

\begin{figure}[h]
   \centering
   \vspace{1.5em}
   \includegraphics[width=\linewidth]{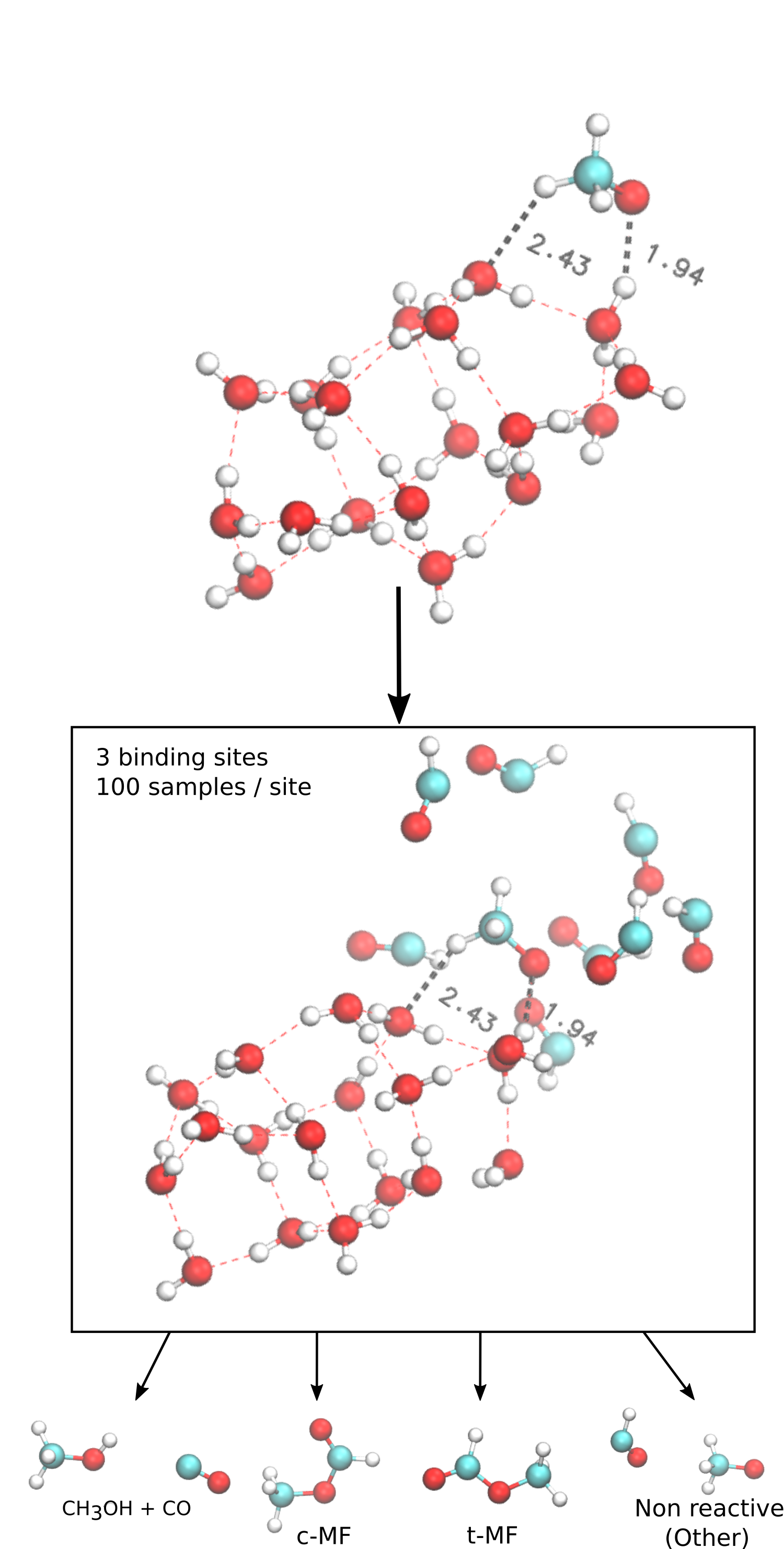} \\
   \caption{Schematic representation of the sampling of reactive outcomes for the \ce{CH3O + HCO} reaction on a 18 \ce{H2O} cluster. Top. Placement of the \ce{CH3O} molecule (Depicted the High site, see text). Center. Simplified representation of the radicals placed around the \ce{CH3O} radical. We depict only 8 starting geometries, for clarity of visualization while in reality they are 100 per binding site. Bottom. Possible outcomes of the simulations. } 
   \label{fig:placement}
\end{figure} 

For each of the initial structures mentioned above, we sequentially placed 100 HCO molecules in random points around the \ce{CH3O} molecule, with the center of mass (CDM) of the HCO molecule placed at 3.5 \AA~+ \{$\max |X|, \max |Y| , \max |Z|$\}, similar to the procedure shown in \citet{Molpeceres2021b}.  $\max |X|, \max |Y| , \max |Z|$ are the absolute values of the maximum Cartesian component of \ce{CH3O} on the ice. The orientation of HCO is randomized. Once the two molecules are placed, the HCO molecule is projected along the vector connecting the CDM of both molecules. This ensures a homogeneous distance between both fragments of $\sim$ 3.5 \AA~. Our routine enforces another constraint where any HCO molecule cannot be at a lower distance to any water molecules in the cluster of 2.5~\AA. In both cases, we refer to interfragment distance, meaning that some atoms might be locally closer or farther away. An example of the sampling is shown in Figure \ref{fig:placement}. We note that we sample both pure Langmuir-Hinshelwood diffusive chemistry and Eley-Rideal trajectories.

Each of the initial configurations is subsequently optimized starting from an open shell singlet wavefunction. Because running 300 geometry optimizations (3 binding sites x 100 samples) is a very demanding task for our computational setup we ran the calculations sequentially. In the first place, we carried out the whole 300 calculations using loose convergence criteria for the optimization. From the results of these calculations, we take the ones that already converged to a reactive result and considered finished. For the rest, we restart the calculations tightening the geometry relaxation convergence criteria, fixing the cluster's atoms degrees of freedom. We repeat this procedure until tight convergence criteria are imposed. The sequential convergence criteria correspond to \texttt{Loose}, \texttt{Normal} and \texttt{Tight} in the \textsc{Orca 5.0.4} package (see Table \ref{tab:opt_criteria}; \citealt{Neese2012, Neese2020}). The trajectories that do not show appreciable reactivity even under stringent criteria are labeled as ``non reactive", in a similar flavour to \cite{lamberts_formation_2019} or \cite{enrique-romero_quantum_2022}. A tiny number (2/300) of trajectories needed to be discarded, due to the break of the open-shell singlet wavefunction during the optimization procedure. These trajectories are simply removed from any posterior analysis.

Once the simulations are concluded we analyze them using statistical approaches. Each outcome is automatically sampled based on three geometrical parameters. The O-H distance is used to probe the formation of \ce{CH3OH}, the C-O distance in MF is used to probe the formation of any isomer of MF, and the dihedral angle $\angle_{\text{COCO}}$ of the participating radicals to sample the isomerism of the product. Any structure not falling into any of above cases is labeled "Non reactive". After considering all possible outcomes, and because our protocol finds a significant number of non-reactive events, we evaluate our results based on a statistical analysis. Each outcome uncertainty is drawn from a binomial distribution of outcomes, calculated using a Jeffreys interval and a confidence level of 95\%. We use the \textsc{statsmodels} Python library for the statistical analysis \citep{seabold2010statsmodels}. Finally, the uncertainties on the ratios for the different outcomes N$_{\ce{CH3OH}}$ /N$_{\text{MF}}$ and N$_{trans-MF}$ /N$_{cis-MF}$ were obtained based on error propagation theory, using as uncertainty for each number the highest of their upper and lower bounds obtained in the binomial analysis mentioned above.

\subsection{Quantum chemical results}
\label{subsec:CompRes}

\begin{figure}[h]
   \centering
   \vspace{1.5em}
   \includegraphics[width=0.5\linewidth]{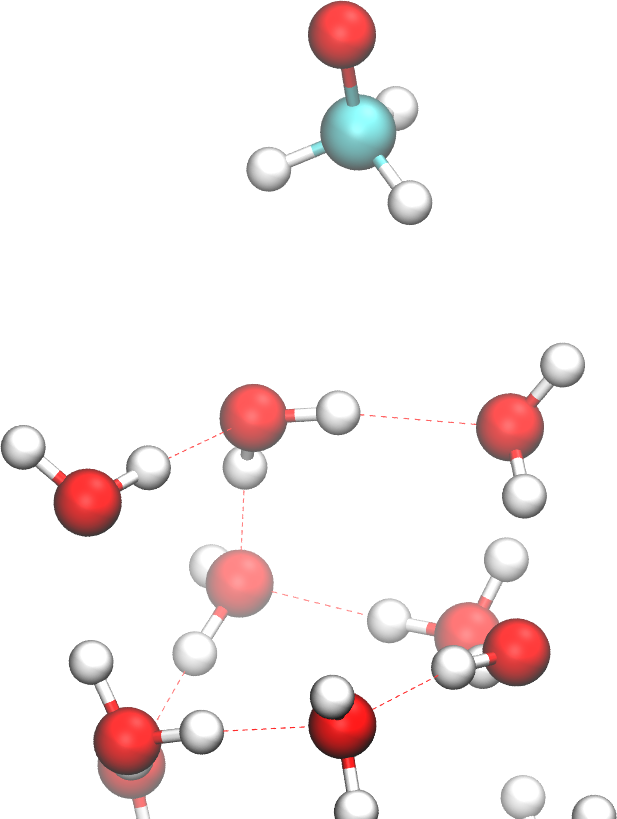} \\
   \caption{Initial placement of the weak binding site.} 
   \label{fig:weak}
\end{figure}

The results of our sampling are reported in Table \ref{tab:values}. An inspection of the table quickly reveals several conclusions regarding the formation of MF, and suggests new avenues to improve our results. It is revealing that non-reactive events tend to dominate the outcomes, with nearly 50 \% of the outcomes in High and Medium binding sites. This number is reduced for the sampling in Low binding sites, owing to the geometry of the \ce{CH3O} radical in this site, where the \ce{-CH3} moiety is more accessible to the attack of \ce{HCO} (See Figure \ref{fig:weak}). Nevertheless, the amount of non-reactive events is, in all cases high. This outcome can be seen as a combination of a physical effect, not all orientations are reactive, as shown in \citet{enrique-romero_quantum_2022, Simons2020, Molpeceres2021b, Molpeceres2022} and a non-physical effect arising from the limited tolerance of the geometry optimizer in static quantum chemical calculations. Improving on the description of branching ratios in radical-radical reactions is in our immediate research plans, through molecular dynamics simulations to palliate the non-physical effect generated by the optimizer. Nevertheless, our current approach allow us to draw sufficient conclusions for the \ce{CH3O + HCO} reaction. 

First commenting on the $N$$_{\ce{CH3OH}}$ /$N$$_{\text{MF}}$ ratio we observe that the formation of methanol, evincing the occurrence of hydrogen abstraction reactions, generally dominates over addition, except for the Medium binding site where the ratio approaches one. Taking the average of $N$$_{\ce{CH3OH}}$ /$N$$_{\text{MF}}$\footnote{Averaging High, Medium and Low binding sites assume equal population of these sites on the surface, which is a rather crude oversimplification. However, we maintain the comparison for illustrative purposes.} with an appropriate treatment of error propagation, we arrive at $\overline{\left( \dfrac{N_{\ce{CH3OH}}}{N_{\text{MF}}} \right)}$ $\sim$ 1.3, showing slightly preferential formation of \ce{CH3OH + CO} in our sampling.

Second, we inspect the ratio $N$$_{\text{trans-MF}}$ / $N$$_{\text{cis-MF}}$ that is more interesting for the purpose of this article. We observe that, while the formation of \textit{cis}-MF greatly dominates in our sampling, we do find a tiny amount of \textit{trans}-MF (1/300 trajectories), which improves the statistical robustness of our analysis. Extracting $\overline{\left( \dfrac{N_{\text{trans-MF}}}{N_{\text{cis-MF}}} \right)}$ we arrive at $\sim$0.013 (or inverting the numerator and denominator $\sim$75). However, we note that in the last column of Table \ref{tab:values} the uncertainty is higher than the derived value. Our results unambiguously indicate that the \ce{CH3O + HCO} pathway is stereoselective, i.e., that the formation of \textit{cis}-MF is markedly dominant (as shown in Figure \ref{f:model} with a thick solid line), although \textit{trans}-MF is also produced as a rarity, which is the most important result of the theoretical part of our study. The stereoespecific formation of a certain isomeric form is in agreement with our previous study for \ce{HC(O)SH} \citep{Molpeceres2021b}, but at odds with our newest work on \ce{C2H4O2} \citep{JC2024}. Overall, it is not possible to rule out a priori which routes are stereospecific on the surface, as these depend on the specific binding mode of the molecule with it. Also, the fact that \textit{trans}-MF is barely formed on the surface indicates that, approximations of our limiting sampling aside, the isomeric ratio may not be fully explained relying on the \ce{CH3O + HCO -> MF} formation route alone. Therefore, as described in the following section, further (mostly gas-phase) pathways will be pivotal in fine tuning the derived \textit{cis}/\textit{trans} ratio.

\section{Discussion: Can stereoselective grain reactions explain the observed isomeric ratio?} 
\label{sec:discussion}

\begin{table*}[htbp]
\centering
\caption{Outcome of the theoretical simulations in different binding sites. } \label{tab:values}
    \begin{tabular}{l|ccccccc}
    \toprule   
	Binding Site & $N$$_{\text{Total}}$ & $N$$_{\ce{CH3OH}}$$^{a}$ &	$N$$_{\text{cis-MF}}$$^{a}$ &	$N$$_{\text{trans-MF}}$$^{a}$	& $N$$_{\text{Non-reactive}}$$^{a}$&  $N$$_{\ce{CH3OH}}$ /$N$$_{\text{MF}}$$^{b,c}$ & $N$$_{\text{trans-MF}}$ / $N$$_{\text{cis-MF}}$$^{b}$ \\
    \bottomrule
    High & 100 & $25^{+9.1}_{-7.7}$ & $19^{+8.5}_{-6.7}$ & $0^{+2.5}_{-0.0}$ & $56^{+9.4}_{-9.8}$ & 1.3 $\pm$ 0.8 & $0^{+0.1}_{-0.0}$  \\
    Medium & 98 & $21^{+8.7}_{-7.1}$ & $27^{+9.1}_{-7.8}$ & $1^{+3.6}_{-0.9}$ & $51^{+9.6}_{-9.6}$ & 0.8 $\pm$ 0.4 & 0.04 $\pm$ 0.1 \\
    Low & 100 & $49^{+9.7}_{-9.7}$ & $28^{+9.3}_{-8.0}$ & $0^{+2.5}_{-0.0}$ & $23^{+8.9}_{-7.4}$ & 1.8 $\pm$ 0.7 & $0^{+0.1}_{-0.0}$ \\
%    ER22 & 98 & $8^{+6.5}_{-4.1}$ & $21^{+8.7}_{-7.1}$ & $10^{+7.0}_{-4.7}$ & $59^{+9.7}_{-9.1}$ &  0.3 $\pm$ 0.2 & 0.50 $\pm$ 0.4 \\
\bottomrule
\end{tabular}
\tablefoot{N: Number of sampled binding sites where the subscript indicates for which species. $^{a}$: Errors obtained from a binomial distribution of outcomes (i.e., Outcome / Total number of samples) assuming a confidence interval of 95\%. $^{b}$: Errors obtained through error propagation taking the highest value of the uncertainty in $N$$_{X}$. For values of 0 in the last column, we have adopted the nomenclature used for the non-symmetric uncertainties. $^{c}$: The $N$$_{\ce{CH3OH}}$ /$N$$_{\text{MF}}$ are only reported for our simulations of reaction 1 while, in reality, observations clearly show a much higher abundance of \ce{CH3OH} over MF in this source \citep{rodriguez-almeida2021a, jimenez-serra2022}. This evinces that the main \ce{CH3OH} formation route is different to \ce{CH3O + HCO -> CH3OH + CO}, e.g., \ce{CO + 4H -> CH3OH} \citep{Watanabe:2002od,jimenez-serra2025}.}
\end{table*}

On the basis of the observed molecular abundances, we derive a \textit{cis}/\textit{trans} isomeric ratio of 72 $\pm$ 16 and 34 $\pm$ 4 toward G+0.693 and L1157-B1, respectively, which can be put together with the previously reported value of $\sim$7 observed toward the envelope of Sgr B2(N) \citep{Neill:2012fr}. At this point, a question naturally arises: are there any common ingredients between the three astronomical sources where \textit{trans}-MF has been detected? As mentioned in Section \ref{sec:obs}, the chemistry of G+0.693 and L1157-B1 is clearly dominated by interstellar shocks. The same applies to the envelope of Sgr B2(N), as supported by the observation of shock-related species such as SiO all over the envelope of Sgr B2 cloud \citep{martin-pintado_sio_1997}, and also the tentative detection of FeO or SiN \citep{Walmsley:2002ud,Schilke2003}. Moreover, these three environments exhibit enhanced cosmic-ray ionization rates (CRIRs) compared to the standard Galactic disk value of $\zeta$=1.3$\times$10$^{-17}$ s$^{-1}$ \citep{Padovani2009}, highlighting values above $\zeta$ $\sim$ 10$^{-15}$ s$^{-1}$ in the Central Molecular Zone (CMZ) \citep{goto2014,Yusef-zadeh2024}. For instance, $\zeta$ $\sim$ 10$^{2}$-10$^{3}$ times higher than the canonical one are needed to reproduce the observations of recently detected ions toward G+0.693 (i.e., PO$^+$ and HOCS$^+$; \citealt{rivilla2022b, Sanz-Novo2024a}), while slightly lower CRIRs of $\sim$2-3$\times$10$^{-16}$ s$^{-1}$ and $\sim$4$\times$10$^{-16}$ s$^{-1}$ have been constrained for L1157-B1 \citep{Podio:2014eq,Luo2024}) and the Sgr B2 envelope \citep{vandertak2006}). These high CRIRs, as we will show below, may also play a crucial role in the intricate balance of formation and destruction of \textit{trans}-MF, ultimately enabling its detection.

\begin{figure}
\centerline{\resizebox{1\hsize}{!}{\includegraphics[angle=0]{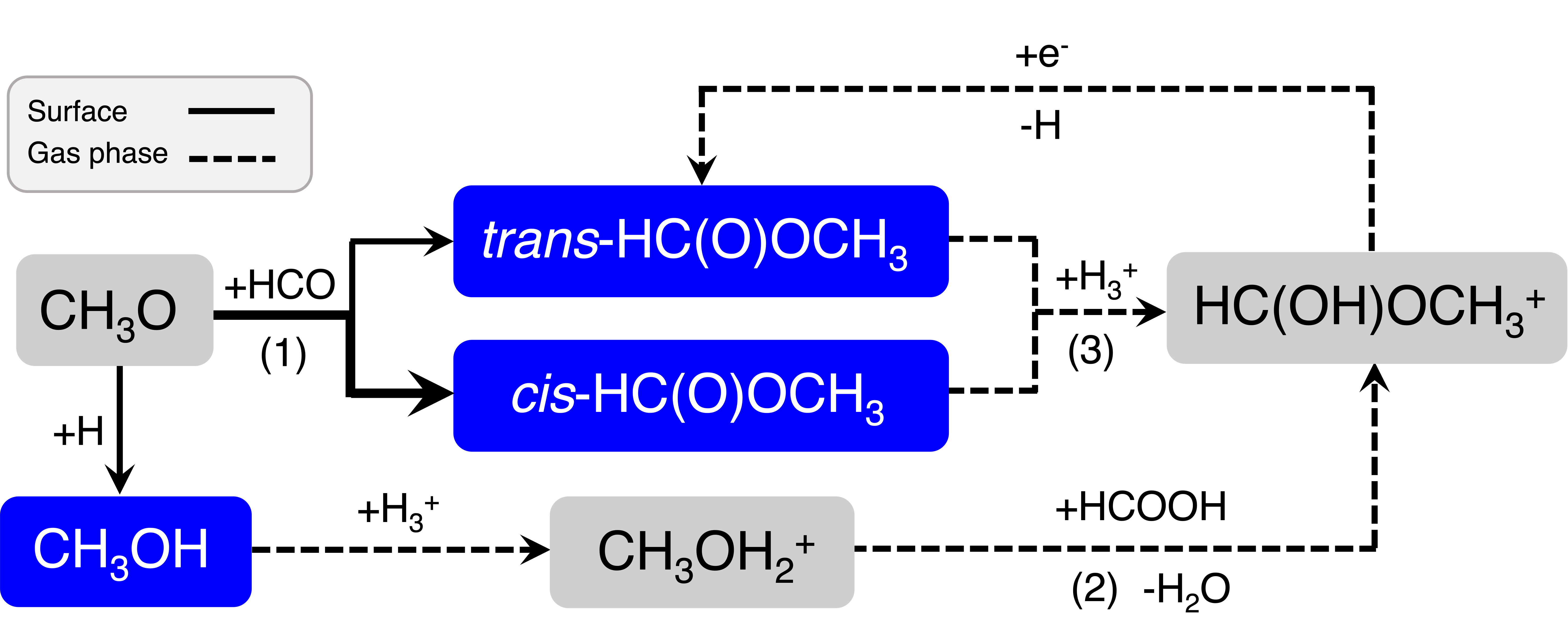}}}
\caption{Suggested stereoselective chemical routes for the formation of MF in the ISM discussed in this work. Molecules that have been identified toward the three target sources are depicted in blue. Surface reactions are shown in black solid lines while gas-phase reactions are depicted with dashed arrows. The thick solid line represents a dominant formation of \textit{cis}-MF.}
\label{f:model}
\end{figure}

The results from our theoretical computations on grains suggest a stereoselective formation of \textit{trans}-MF via the \ce{CH3O + HCO} route (reaction \ref{eq:formation} shown in Figure \ref{f:model}). We derive a \textit{trans}/\textit{cis} ratio comprised between 0.0 and 0.14 -considering the uncertainties in the computations-, which is in reasonable agreement with those observed, ranging between 0.014--0.1 for the three target sources. Strikingly, our approximated \textit{cis}/\textit{trans} theoretical value of $\sim$75 (after inverting the numerator and denominator) nicely matches the ratio found for G+0.693 of $\sim$72, which suggests that \textit{trans}-MF is selectively formed on icy grain surfaces. In this context, it is thought that the energetic conditions of G+0.693 promote non-thermal radical chemistry on the ices, that are easily released to the gas phase through shocks. Therefore, the gas in G+0.693, but also that in L1157-B1 and the envelope of Sgr B2(N), serves as an excellent tracer for the chemistry on the ices. Also, we note that the transient and highly reactive \ce{CH3O} radical is often challenging to detect in radioastronomical observations. When observed, its abundance is significantly low —even lower than that of MF \citep{Jimenez-Serra16}- as it can rapidly react to form more complex products. Thus, although \ce{CH3O} has not yet been detected toward the target regions, we anticipate an efficient production as a result of UV secondary photons induced by cosmic rays interacting with the surface of dust grains, which photodissociate \ce{CH3OH}.

Although the variability of the isomeric ratio found across the three sources could be directly inherited from grain chemistry, additional gas-phase events contributing to a selective formation of \textit{trans}-MF cannot be ruled out a priori. Once \textit{trans}-MF reaches the gas phase, the large computed barrier for the relaxation process from \textit{trans}-MF to \textit{cis}-MF ( $\sim$13.8 kcal mol${-1}$ or 6944 K \citealt{Senent2005,Neill:2012fr}) is likely to prevent the conformational isomerization from occurring under ISM conditions. Ground-state quantum tunneling effects are also likely greatly diminished, in contrast to other examples occurring through a single H-atom migration \citep{garciadelaconcepcion2021,garciadelaconcepcion2022}, due to the much heavier methyl group. Such hypothetical scenario would result in an exceedingly high \textit{cis}:\textit{trans} ratio under thermal equilibrium conditions (e.g., approximately 10$^{13}$:1 at 100 K), %which is far from the observed values in any of the three astronomical sources. 
which would mean that the \textit{trans} conformer should not be detectable in the ISM.

In this scenario, we need to explore alternative gas-phase processes that have already been proposed to favor the production of \textit{trans}-MF over \textit{cis}-MF  \citep{Neill:2011pd,Neill:2012fr,Cole2012}. One of the most promising pathways is the following ion-molecule reaction \ref{eq:protformic} (also depicted in Figure \ref{f:model}) between protonated methanol (\ce{CH3OH2+}) and formic acid (HCOOH):

\begin{equation}
\mathrm{\ce{CH3OH2+}} + \mathrm{\ce{HCOOH}} \rightarrow \rm \mathrm{\ce{HC(OH)OCH3+}} + \mathrm{\ce{H2O}} 
\label{eq:protformic}
\end{equation}

This exothermic and barrierless reaction was studied theoretically by \citet{Neill:2011pd} and in the laboratory by \citet{Cole2012}, and it was also included in the models of \citet{laas2011}. The route involves two distinct transition state geometries that lead to different conformations of protonated MF, depending on the \textit{cis} or \textit{trans} arrangements of the C–O–C–O dihedral angle. Interestingly, the formation of \textit{trans}-MF proceeds without an activation barrier, while the synthesis of the \textit{cis} conformer implies a net barrier of 10 kJ mol $^{-1}$. Thus, reaction \ref{eq:protformic} appears to be stereoselective and may preferentially yield \textit{trans}-MF through dissociative recombination. This route only requires that both reactants, HCOOH an \ce{CH3OH}, are produced on grains and once desorbed into the gas phase, \ce{CH3OH} can be protonated by the reaction with \ce{H3+} \citep{Fiaux1976}, followed by the formation of \textit{trans}-MF. Indeed, HCOOH is abundant in the three studied sources \citep{Winnewisser:1975we,lefloch_l1157-b1_2017,Sanz-Novo2023}, and we expect that \ce{CH3OH2+} is also present, given the high abundance of \ce{CH3OH}, and the aforementioned high CRIRs found in these environments. Thus, this route may significantly contribute to the formation and subsequent detection of \textit{trans}-MF toward the target regions.

In sharp contrast, an additional acid-catalyzed Fischer esterification reaction between neutral methanol and protonated formic acid \citep{Neill:2011pd} (i.e., \ce{CH3OH} + \ce{HC(OH)2+} $\rightarrow$ \ce{HC(OH)OCH3+} + \ce{H2O}) was quickly ruled out. This stems from the formation of both \textit{cis} and \textit{trans}-MF, which involves significant activation barriers (17 kJ mol$^{-1}$ and 21 kJ mol$^{-1}$, translated in 2045 K and 2526 K, respectively), and therefore is not viable under the relatively low gas kinetic temperatures of the studied interstellar sources.

\citet{Neill:2012fr} also proposed an alternative mechanism for the conformational isomerization involving an indirect process. In this scenario, MF could first be protonated by a prevalent molecular ion (reaction 3, shown in Figure \ref{f:model}), such as \ce{H3+}, and then undergo dissociative recombination. Both of these steps are highly exothermic, according to gas-phase proton affinities \citep{Hunter1998}, with energy releases that may surpass the isomerization barriers between the neutral \textit{cis} and \textit{trans} conformers of MF. Thus, since the barriers for isomerization in the protonated form are estimated to be similar, the highly energetic molecule might undergo isomerization during the relaxation phase of either reaction step. As a result, this process could create a cyclical effect, leading to a non-thermal conformational abundance ratio, which may also significantly contribute to the formation of \textit{trans}-MF. This route is in essence similar to the recently proposed \textit{sequential acid-base} mechanism used to explain formic acid (\ce{HCOOH}) isomerism \citep{garcia_de_la_concepcion_sequential_2023} suggesting that this might be a common mechanism to explain selective isomerism. 

The last of the known mechanisms that can be considered to produce an isomeric excess is the interconversion on grain surfaces through a two step hydrogen interconversion (H-abstraction + Hydrogen addition). It is known that this mechanism produces an excess of the most stable conformer of HCOOH \citep{Molpeceres2022}. However, in the case of MF we do not expect this mechanism to play any role since the isomeric form is determined by the C-O-C-O dihedral moiety, so there is no chance to initiate the abstraction/addition cycle in the first place.

Additionally, regarding possible stereoselective destruction mechanisms, it is worth noticing that if the relative dipole principle (RDP; \citealt{Shingledecker2020}) holds for this family of conformers, the higher dipole moment of \textit{trans}-MF ($\mu$$_a$ = 4.2 D; \citealt{Neill:2012fr}) will make destruction rates in the gas phase faster than in the case of \textit{cis}-MF ($\mu$$_a$ = 1.6 D; \citealt{Curl1959}), further decreasing the \textit{trans}/\textit{cis} ratio. Therefore, this general "rule of thumb" could also contribute to fine tuning the final observed isomeric ratio, making the detection of \textit{trans}-MF even more challenging, should this mechanism prove efficient.

Overall, combining our quantum chemical results with our knowledge of the objects hosting \textit{trans}-MF, we conclude that stereoselective grain surface routes could qualitatively predict the observations. However, thorough chemical modeling that includes additional gas-phase formation (and destruction) pathways (see e.g., \citealt{Cole2012,Neill:2012fr, Shingledecker2020,garcia_de_la_concepcion_sequential_2023}), will be pivotal to determine whether the \ce{CH3O + HCO -> MF} reaction alone can produce enough \textit{trans}-MF to match the observed isomeric ratio, as well as to establish the dominant formation pathway, if one exists.

\section{Conclusions}
\label{s:conclusions}

In this study, we presented new detections of the high-energy conformer of methyl formate (\textit{trans}-MF) toward two shocked-dominated regions: G+0.693-0.027 and L1157-B1, providing definite observational evidence of its presence in the ISM. Additionally, aiming to shed light on the conformational isomeric ratio of MF, we compared the derived observational ratios with predictions based on new grain-surface chemistry computations. The main conclusions of this work are the following:

\begin{itemize}
    \item We have conclusively confirmed the interstellar detection of \textit{trans}-MF, which was previously tentatively observed in the envelope of Sgr B2(N) as reported by \citet{Neill:2012fr}, through its identification in two shock-dominated regions: G+0.693 and L1157-B1. In both sources, we have identified numerous unblended $^a$$R$-branch $K$$_a$ = 0, 1 transitions belonging to the $A$-symmetry substate of \textit{trans}-MF. Many of these lines have been observed for the first time directly in the radio astronomical data and still remain unmeasured in the laboratory. We derived a molecular column density for \textit{trans}-MF of $N$(G+0.693) = (8.2 $\pm$ 0.4) $\times$10$^{12}$ cm$^{-2}$ and $N$(L1157-B1) = (1.2 $\pm$ 0.2) $\times$10$^{12}$ cm$^{-2}$, yielding a molecular abundance with respect to H$_2$ of $\sim$6 $\times$ 10$^{-11}$ and $\sim$6 $\times$ 10$^{-10}$, respectively, for each source. 

     \item Based on the derived abundances, we obtain a \textit{cis}/\textit{trans} isomeric ratio of $\sim$72 and $\sim$34 toward G+0.693 and L1157-B1, respectively, which are $\sim$7 and 3 times larger than that found in the Sgr B2(N) region and also quite far from the expected thermodynamic ratio (i.e., approximately 10$^{13}$:1 at 100 K).

    \item Our theoretical results suggest that grain surface pathways can qualitatively explain the observed non-thermal \textit{cis}/\textit{trans} abundance ratio, pinpointing to a stereoespecific formation of \textit{trans}-MF via the \ce{CH3O + HCO} route (reaction \ref{eq:formation}). Particularly, we found a nice match between the predicted ratio ($\sim$75) and the observed value toward G+0.693 ($\sim$72), but with a great uncertainty. Nevertheless, additional stereoespecific gas-phase routes (i.e., reactions \ref{eq:protformic}) and 3) may also play a crucial role in the shaping the formation of an excess of \textit{trans}-MF, ultimately enabling its detection. Thus, detailed modeling of the chemistry of \textit{trans}-MF will be key to disclose its primary formation pathway, should it exist.
    
    \item Future observational effort may focus on the search for \textit{trans}-MF toward sources where the most stable \textit{cis} conformer is very abundant and, ideally, environments dominated by shocks and high CRIRs, which appear as two key players driving the detection of the high-energy \textit{trans}-MF conformer. However, additional laboratory measurements will be needed first to enable reliable searches if the target source exhibits very narrow linewidths (see e.g., the QUIJOTE line survey of TMC-1, with FWHM= 0.5$-$0.6 km s$^{-1}$; \citealt{Cernicharo21}).

    \item Finally, we have shown that the exploration of stereoisomerism itself serves as: i) A powerful tool to distinguish between different chemical routes of molecules as molecular complexity increases. ii) A driving motif for the detection of new species in the ISM. In some cases, the most stable conformers are nearly 'invisible' to radioastronomical observations whereas higher-energy species exhibit a sizable dipole moment \citep{Sanz-Novo2023}. Consequently, it will be key to characterize in detail the spectral features and formation mechanisms of not only the most stable conformers, but all the structures that are present in the conformational panorama of the molecule, some of which are potentially detectable in the ISM.

\end{itemize}

\begin{acknowledgements}

We are grateful to the IRAM 30$\,$m and Yebes 40$\,$m telescopes staff for their help during the different observing runs, highlighting project 21A014 (PI: Rivilla), project 018-19 (PI: Rivilla) and project 123-22 (PI: Jim\'enez-Serra). The 40$\,$m radio telescope at Yebes Observatory is operated by the Spanish Geographic Institute (IGN, Ministerio de Transportes, Movilidad y Agenda Urbana). IRAM is supported by INSU/CNRS (France), MPG (Germany) and IGN (Spain). We are grateful for the dataset of the ASAI IRAM-30m Large Program (https://www.iram-institute.org/EN/content-page-344-7-158-240-344-0.html) provided in Zenodo within the framework of ACO (AstroChemical Origins): (H2020 MSCA ITN, GA:811312), and for making it available to the community. M. S.-N. acknowledges a Juan de la Cierva Postdoctoral Fellow proyect JDC2022-048934-I, funded by the Spanish Ministry of Science, Innovation and Universities/State Agency of Research MICIU/AEI/10.13039/501100011033 and by the European Union “NextGenerationEU”/PRTR”. V. M. R. acknowledges support from project number RYC2020-029387-I funded by MCIN/AEI/10.13039/501100011033 and by "ESF, Investing in your future", from the Consejo Superior de Investigaciones Cient{\'i}ficas (CSIC) and the Centro de Astrobiolog{\'i}a (CAB) through the project 20225AT015 (Proyectos intramurales especiales del CSIC), and from the grant CNS2023-144464 funded by MICIU/AEI/10.13039/501100011033 and by “European Union NextGenerationEU/PRTR”. I. J.-S., V. M. R and M. S.-N, acknowledge funding from grant No. PID2022-136814NB-I00 funded by MICIU/AEI/10.13039/501100011033 and by ERDF, UE. I.J.-S. also acknowledges support by ERC grant OPENS, GA No. 101125858, funded by the European Union. Views and opinions expressed are however those of the author(s) only and do not necessarily reflect those of the European Union or the European Research Council Executive Agency. Neither the European Union nor the granting authority can be held responsible for them. G.M acknowledges the support of the grant RYC2022-035442-I funded by MCIN/AEI/10.13039/501100011033 and ESF+. G.M. also received support from project 20245AT016 (Proyectos Intramurales CSIC). We acknowledge the computational resources provided by bwHPC and the German Research Foundation (DFG) through grant no INST 40/575-1 FUGG (JUSTUS 2 cluster), the DRAGO computer cluster managed by SGAI-CSIC, and the Galician Supercomputing Center (CESGA). The supercomputer FinisTerrae III and its permanent data storage system have been funded by the Spanish Ministry of Science and Innovation, the Galician Government and the European Regional Development Fund (ERDF).

\end{acknowledgements}

\bibliography{bibliography,rivilla}
\bibliographystyle{aasjournal}

\begin{appendix}
\label{appendix}

\section{Complementary tables}

In Table \ref{t:mwtable1}, we report the experimental constants for the ground state of \textit{trans}-MF ($A$-species) obtained from a separated effective fit to a semi-rigid rotor Hamiltonian ($A$-Reduction, I$^{r}$-Representation), while in Table \ref{t:pfun} we provide the corresponding rotational ($Q_r$) partition function, needed to obtain reliable line intensities. We used SPCAT \citep{Pickett1991} to compute the values of $Q_r$ by direct summation of the ground state energy levels up to $J$ = 79 as in \citealt{Carvajal2019} for the \textit{cis} conformer. These values are provided for the conventional temperatures as implemented in the Cologne Database for Molecular Spectroscopy (CDMS; \citealt{Muller2005, endres2016}) and JPL database \citep{Pickett:1998cp}, and agree well with the values of $Q_r$ = 4.67 $T_{\rm ex}$$^{1.5}$ used in \citet{Neill:2012fr} for the $A$ species. Also, since the excitation temperature of the molecules in G+0.693 are low ($T_{\rm ex}$\,=\,5$-$20~K), we have not used in this case the vibrational contribution of the partition function ($Q_v$). We then prepared a new catalog for the $A$-symmetry substate of \textit{trans}-MF in the common JPL SPFIT/SPCAT format, which was used to carry out the interstellar search. Table \ref{tab:opt_criteria} provides the numerical values for the different \textbf{\textsc{Orca}} convergence criteria used in this work. Finally, the list of observed transitions for \textit{cis}-MF detected toward G+0.693 and L1157-B1 are given in Tables \ref{tab:transitionscisL1157} and \ref{tab:transitionscisG0693}.

\begin{table}[!ht]
\begin{center}
\caption{Experimental spectroscopic parameters for the $A$-symmetry substate of \textit{trans}-MF ($A$-Reduction, I$^{r}$-Representation).}
\label{t:mwtable1}
\vspace*{0.0ex}
\begin{tabular}{lllll}
\hline\hline
\multicolumn{1}{c}{Parameters} & \multicolumn{1}{c}{\textit{trans}-MF}  \\ 
\hline
$A^{(a)}$\small (MHz) & 62897.147 $\pm$ 0.004$^{(e)}$ \\
$B$ \small (MHz) & 4725.7172 $\pm$ 0.0004  \\ 
$C$ \small (MHz) & 4398.5091 $\pm$ 0.0004  \\ 
|$\mu_a$| / |$\mu_b$| / |$\mu_c$|$^{(b)}$\small (D) & 4.2 / 2.5 / 0.0  \\ 
$\Delta_J$ \small (kHz) & 1.151 $\pm$ 0.009 \\ 
$\Delta_{JK}$ \small (kHz) & -74.6 $\pm$ 0.2  \\ 
$\delta_J$ \small (kHz) & 0.1100 $\pm$ 0.0005  \\ 
$N$. $A$ lines \small  & 24  \\ 
$\sigma^{(c)}$\small (kHz) & 20 \\ 
$\sigma_{w}^{(d)}$\small & 0.59 \\ 
\hline 
\end{tabular}
\end{center}
\vspace*{-1.0ex}
\tablefoot{\tablefootmark{(a)}$A$, $B$, and $C$ represent the rotational constants (in MHz). \tablefootmark{(b)}|$\mu_a$|, |$\mu_b$|, |$\mu_c$| are the absolute values of the electric dipole moment components (in D), obtained theoretically in \citet{Neill:2012fr}. \tablefootmark{(c)} Root mean square (RMS) deviation of the fit. \tablefootmark{(d)} Dimensionless \emph{rms}, defined as $\sigma_{w} = \sqrt{\frac{\sum_i\left(\delta_i/err_i\right)^2}{N_A}}$, where the $\delta$'s are the residuals weighted by the experimental uncertainty and ${N_A}$ the total number of measured transitions analyzed belonging to the $A$-symmetry substate.}
\end{table}

\begin{table}
\begin{center}
\caption{Rotational ($Q$$_r$) partition function of the $A$-symmetry substate of \textit{trans}-MF.}
\label{t:pfun}
\begin{tabular}{ccc}
\hline
\multicolumn{1}{c}{Temperature}{\small (K)} & \multicolumn{1}{c}{$Q$$_r$} & \multicolumn{1}{c}{log($Q$$_r$)} \\  
\hline
9.375 &	134.8452 & 2.1298 \\
18.75 &	379.9630 & 2.5797 \\
37.50 &	1072.666 & 3.0305 \\
75.00 &	3031.037 & 3.4816 \\
150.0 &	8567.828 & 3.9329 \\
225.0 &	15705.37 & 4.1960 \\
300.0 &	23988.53 & 4.3800 \\
\hline 
\end{tabular}
\end{center}
\vspace*{-1.0ex}
%\vspace{-2mm}
\end{table}

\begin{table}[h]
    \centering
    \tabcolsep 2.5pt
    \caption{\textsc{Orca} convergence criteria used in this work.}
    \begin{tabular}{lccc}
        \toprule
        Convergence criteria keyword & LooseOpt & NormalOpt & TightOpt \\
        \midrule
        Energy (E$_h$) & $3\times10^{-5}$ & $5\times10^{-6}$ & $1\times10^{-6}$ \\
        RMS Gradient (E$_h$ Bohr$^{-1}$) $^{(a)}$ & $5\times10^{-4}$ & $1\times10^{-4}$ & $3\times10^{-5}$ \\
        MAX Gradient (E$_h$ Bohr$^{-1}$) & $2\times10^{-3}$ & $3\times10^{-4}$ & $1\times10^{-4}$ \\
        RMS Step (Bohr) & $7\times10^{-3}$ & $2\times10^{-3}$ & $6\times10^{-4}$ \\
        MAX Step (Bohr) & $1\times10^{-2}$ & $4\times10^{-3}$ & $1\times10^{-3}$ \\
        \bottomrule
    \end{tabular}
    \label{tab:opt_criteria}
    \vspace*{-1.0ex}
    \tablefoot{\tablefootmark{(a)}RMS refers to the Root Mean Square Deviation of the Gradient.}
\end{table}

\begin{table*}
\centering
\tabcolsep 4.5pt
\caption{Sample list of the observed transitions of \textit{cis}-MF detected toward G+0.693. We provide the transitions frequencies, quantum numbers, base 10 logarithm of the integrated intensity at 300 K (log $I$), and the values of the upper levels of each transition ($E_{\rm u}$).}
\begin{tabular}{ c c c c l}
\hline
 Frequency & Transition  & log $I$ & $E_{\rm u}$  & Blending$^a$ \\
 (GHz) & ($J_{K_a,K_c}$)  &   (nm$^2$ MHz)  & (K) &   \\
\hline
45.395795 & 4$_{1,4}$ - 3$_{1,3}$ ($E$)  & -5.8712  &  6.1 & Unblended$^{*}$   \\  %OK
45.397380 & 4$_{1,4}$ - 3$_{0,3}$ ($A$)  & -5.8710  &  6.1 & Unblended$^{*}$  \\  %OK
47.534093 & 4$_{0,4}$ - 3$_{0,3}$ ($E$)  & -5.8067  &  5.7 & Unblended$^{*}$  \\  %OK
47.536915 & 4$_{0,4}$ - 3$_{0,3}$ ($A$)  & -5.8065  &  5.7 & Unblended$^{*}$  \\  %OK
72.680833 & 6$_{2,5}$ - 5$_{2,4}$ ($E$)  & -5.3215  &  14.8 & Unblended$^{*}$  \\  %OK
72.685565 & 6$_{2,5}$ - 5$_{2,4}$ ($A$)  & -5.3213  &  14.8 & Unblended$^{*}$  \\  %OK
73.885081 & 6$_{3,4}$ - 5$_{3,3}$ ($A$)  & -5.3846  &  18.2 & Unblended$^{*}$  \\  %OK
73.905915 & 6$_{3,4}$ - 5$_{3,3}$ ($E$)  & -5.4082  &  18.3 & Unblended$^{*}$  \\  %OK
74.263465 & 6$_{3,3}$ - 5$_{3,2}$ ($E$)  & -5.4040  &  18.3 & Unblended$^{*}$  \\  %OK
74.296741 & 6$_{3,3}$ - 5$_{3,2}$ ($A$)  & -5.3798  &  18.3 & Unblended$^{*}$  \\  %OK
76.701827 & 6$_{2,4}$ - 5$_{2,3}$ ($E$)  & -5.2734  &  15.3 & Unblended$^{*}$  \\  %OK
76.711146 & 6$_{2,4}$ - 5$_{2,3}$ ($A$)  & -5.2732  &  15.2 & Unblended$^{*}$  \\  %OK
76.796033 & 6$_{1,5}$ - 5$_{1,4}$ ($E$)  & -5.2347  &  13.6 & Blended with U  \\  %OK
76.803994 & 6$_{1,5}$ - 5$_{1,4}$ ($A$)  & -5.2345  &  13.6 & Blended with $^{13}$CH$_3$CHO and U  \\  %OK
78.479392 & 7$_{1,7}$ - 6$_{1,6}$ ($E$)  & -5.1476  &  15.8 & Unblended$^{*}$  \\  %OK 
78.481388 & 7$_{1,7}$ - 6$_{1,6}$ ($A$)  & -5.1465  &  15.7 & Unblended$^{*}$  \\  %OK 
79.781707 & 7$_{0,7}$ - 6$_{0,6}$ ($E$)  & -5.1302  &  15.6 & Unblended$^{*}$  \\  %OK  
79.783887 & 7$_{0,7}$ - 6$_{0,6}$ ($A$)  & -5.1301  &  15.6 & Unblended$^{*}$  \\   %OK
84.449169 & 7$_{2,6}$ - 6$_{2,5}$ ($E$)  & -5.1160  &  18.9 & Unblended$^{*}$  \\   %OK  
84.454754 & 7$_{2,6}$ - 6$_{2,5}$ ($A$)  & -5.1158  &  18.8 & Unblended$^{*}$  \\   %OK 
88.843187 & 7$_{1,6}$ - 6$_{1,5}$ ($E$)  & -5.0452  &  17.8 & Unblended$^{*}$  \\  %OK  
88.851607 & 7$_{1,6}$ - 6$_{1,5}$ ($A$)  & -5.0450  &  17.8 & Unblended$^{*}$  \\  %OK  
89.314657 & 8$_{1,8}$ - 7$_{1,7}$ ($E$)  & -4.9813  &  20.0 & Blended with $^{13}$CH3CN  \\  %OK 
89.316642 & 8$_{1,8}$ - 7$_{1,7}$ ($A$)  & -4.9811  &  20.0 & Blended with $^{13}$CH3CN  \\  %OK  
90.145723 & 7$_{2,5}$ - 6$_{2,4}$ ($E$)  & -5.0573  &  19.5 & Unnblended  \\   %OK 
90.156473 & 7$_{2,5}$ - 6$_{2,4}$ ($A$)  & -5.0570  &  19.5 & Blended with $t$-HCOOH  \\  %OK 
90.227659 & 8$_{0,8}$ - 7$_{0,7}$ ($E$)  & -4.9710  &  19.9 & Unblended$^{*}$  \\  %OK
90.229624 & 8$_{0,8}$ - 7$_{0,7}$ ($A$)  & -5.9708  &  19.9 & Unblended$^{*}$  \\   %OK
96.070725 & 8$_{2,7}$ - 7$_{2,6}$ ($E$)  & -4.9438  &  23.4 & Unblended$^{*}$  \\ %OK 
96.076845 & 8$_{2,7}$ - 7$_{2,6}$ ($A$)  & -4.9436  &  23.4 & Unblended$^{*}$  \\ %OK  
98.606856 & 8$_{3,6}$ - 7$_{3,5}$ ($E$)  & -4.9628  &  27.1 & Blended with C$_2$H$_5$CN  \\  %OK 
98.611163 & 8$_{3,6}$ - 7$_{3,5}$ ($A$)  & -4.9621  &  27.1 & Blended with C$_2$H$_5$CN   \\  %OK 
100.294604 & 8$_{3,5}$ - 7$_{3,4}$ ($E$)  & -4.9479  &  27.2 & Unblended  \\ %OK 
100.308179 & 8$_{3,5}$ - 7$_{3,4}$ ($A$)  & -4.9472  &  27.2 & Blended with H$_2$NCO$^+$  \\ 
100.482241 & 8$_{1,7}$ - 7$_{1,6}$ ($E$)  & -4.8865  &  22.6 & Unblended$^{*}$  \\ %OK
100.490682 & 8$_{1,7}$ - 7$_{1,6}$ ($A$)  & -4.8863  &  22.6 & Blended with N$_2$O \\ %OK
100.681545 & 9$_{0,9}$ - 8$_{0,8}$ ($E$)  & -4.8307  &  24.7 & Unblended$^{*}$  \\   %OK
100.683368 & 9$_{0,9}$ - 8$_{0,8}$ ($A$)  & -4.8306  &  24.7 & Unblended$^{*}$  \\   %OK
103.466572 & 8$_{2,6}$ - 7$_{2,5}$ ($E$)  & -4.8771  &  24.5 & Unblended  \\ %OK
103.478663 & 8$_{2,6}$ 7$_{2,5}$ ($A$)  & -4.8769  &  24.5 & Unblended  \\ %OK
107.537258 & 9$_{2,8}$ - 8$_{2,7}$ ($E$)  & -4.7963  &  28.6 & Unblended$^{*}$  \\ %OK
107.543711 & 9$_{2,8}$ - 8$_{2,7}$ ($A$)  & -4.7961  &  28.6 & Unblended$^{*}$  \\ %OK
110.788664 &10$_{1,10}$ - 9$_{1,9}$ ($E$)  & -4.7900  &  30.1 & Blended with CH$_3$CHNH  \\ %OK
110.790526 &10$_{1,10}$ - 9$_{1,9}$ ($A$)  & -4.7088  &  30.1 & Blended with CH$_3$CHNH  \\ %OK
110.879766 & 9$_{3,7}$ - 8$_{3,6}$ ($E$)  & -4.8021  &  32.4 & Unblended$^{*}$  \\ %OK
110.887092 & 9$_{3,7}$ - 8$_{3,6}$ ($A$)  & -4.8018  &  32.3 & Unblended$^{*}$  \\ %OK
111.169903 &10$_{0,10}$ - 9$_{0,9}$ ($E$)  & -4.7056  &  30.0 & Unblended$^{*}$  \\ %OK
111.171634 &10$_{0,10}$ - 9$_{0,9}$ ($A$)  & -4.7055  &  30.0 & Unblended$^{*}$  \\ %OK
111.674131 & 9$_{1,8}$ - 8$_{1,7}$ ($E$)  & -4.7514  &  27.9 & Blended with CH$_3$CONH$_2$ and $^{13}$CH$_3$CHO  \\%OK
111.682189 & 9$_{1,8}$ - 8$_{1,7}$ ($A$)  & -4.7512  &  27.9 & Unblended  \\ %OK
113.743107 & 9$_{3,6}$ - 8$_{3,5}$ ($E$)  & -4.7796  &  32.6 & Blended with CH$_3$NCO  \\  %OK
113.756610 & 9$_{3,6}$ - 8$_{3,5}$ ($A$)  & -4.7793  &  32.6 & Unblended  \\  %OK
129.296357 &10$_{2,8}$ - 9$_{2,7}$ ($E$)  & -4.5937  &  36.2 & Unblended  \\  %OK
129.310166 &10$_{2,8}$ - 9$_{2,7}$ ($A$)  & -4.5935  &  36.2 & Unblended  \\  %OK
\hline 
\end{tabular}
\label{tab:transitionscisG0693}
{\\ (a) ``U" refers to blending with emission from an unknown (not identified) species; transitions with the $^{*}$ symbol are not blended with emission from other species, but (auto)blended with another transition of \textit{cis}-MF.
}
\end{table*}

\begin{table*}
\centering
\tabcolsep 4.5pt
\caption{List of observed transitions of \textit{cis}-MF detected toward L1157-B1. We provide the transitions frequencies, quantum numbers, base 10 logarithm of the integrated intensity at 300 K (log $I$), and the values of the upper levels of each transition ($E_{\rm u}$).}
\begin{tabular}{ c c c c l}
\hline
 Frequency & Transition  & log $I$ & $E_{\rm u}$  & Blending$^a$ \\
 (GHz) & ($J_{K_a,K_c}$)  &   (nm$^2$ MHz)  & (K) &   \\
\hline
76.796033 & 6$_{1,5}$ - 5$_{1,4}$ ($E$)  & -5.2347  &  13.6 & Unblended  \\ %OK  
76.803994 & 6$_{1,5}$ - 5$_{1,4}$ ($A$)  & -5.2345  &  13.6 & Unblended  \\ %OK   
79.781707 & 7$_{0,7}$ - 6$_{0,6}$ ($E$)  & -5.1302  &  15.6 & Unblended  \\ %OK  
79.783887 & 7$_{0,7}$ - 6$_{0,6}$ ($A$)  & -5.1301  &  15.6 & Unblended  \\  %OK
84.449169 & 7$_{2,6}$ - 6$_{2,5}$ ($E$)  & -5.1160  &  18.9 & Unblended  \\  %OK
84.454754 & 7$_{2,6}$ - 6$_{2,5}$ ($A$)  & -5.1158  &  18.9 & Unblended  \\  %OK
88.843187 & 7$_{1,6}$ - 6$_{1,5}$ ($E$)  & -5.0452  &  17.8 & Unblended  \\  %OK
88.851607 & 7$_{1,6}$ - 6$_{1,5}$ ($A$)  & -5.0450  &  17.8 & Unblended  \\  %OK
89.314657 & 8$_{1,8}$ - 7$_{1,7}$ ($E$)  & -4.9813  &  20.0 & Unblended$^{*}$  \\  %OK
89.316642 & 8$_{1,8}$ - 7$_{1,7}$ ($A$)  & -4.9811  &  20.0 & Unblended$^{*}$  \\  %OK
90.145723 & 7$_{2,5}$ - 6$_{2,4}$ ($E$)  & -5.0573  &  19.5 & Unblended  \\  %OK
90.156473 & 7$_{2,5}$ - 6$_{2,4}$ ($A$)  & -5.0570  &  19.5 & Unblended  \\  %OK
90.227659 & 8$_{0,8}$ - 7$_{0,7}$ ($E$)  & -4.9710  &  19.9 & Unblended$^{*}$  \\  %OK
90.229624 & 8$_{0,8}$ - 7$_{0,7}$ ($A$)  & -5.9708  &  19.9 & Unblended$^{*}$  \\   %OK
96.070725 & 8$_{2,7}$ - 7$_{2,6}$ ($E$)  & -4.9438  &  23.4 & Unblended  \\  %OK
96.076845 & 8$_{2,7}$ - 7$_{2,6}$ ($A$)  & -4.9436  &  23.4 & Unblended  \\  %OK
100.482241 & 8$_{1,7}$ - 7$_{1,6}$ ($E$)  & -4.8865  &  22.6 & Unblended  \\ %OK
100.490682 & 8$_{1,7}$ - 7$_{1,6}$ ($A$)  & -4.8863  &  22.6 & Unblended  \\ %OK
100.681545 & 9$_{0,9}$ - 8$_{0,8}$ ($E$)  & -4.8307  &  24.7 & Unblended$^{*}$  \\ %OK
100.683368 & 9$_{0,9}$ - 8$_{0,8}$ ($A$)  & -4.8306  &  24.7 & Unblended$^{*}$  \\ %OK
103.466572 & 8$_{2,6}$ - 7$_{2,5}$ ($E$)  & -4.8771  &  24.5 & Unblended  \\ %OK
103.478663 & 8$_{2,6}$ - 7$_{2,5}$ ($A$)  & -4.8769  &  24.5 & Unblended  \\ %OK
107.537258 & 9$_{2,8}$ - 8$_{2,7}$ ($E$)  & -4.7963  &  28.6 & Unblended  \\ %OK
107.543711 & 9$_{2,8}$ - 8$_{2,7}$ ($A$)  & -4.7961  &  28.6 & Unblended  \\ %OK
110.788664 & 10$_{1,10}$ - 9$_{1,9}$ ($E$)  & -4.7900  &  30.1 & Unblended$^{*}$  \\ %OK
110.790526 & 10$_{1,10}$ - 9$_{1,9}$ ($A$)  & -4.7088  &  30.0 & Unblended$^{*}$  \\ %OK
111.674131 & 9$_{1,8}$ - 8$_{1,7}$ ($E$)  & -4.7514  &  27.9 & Unblended  \\ %OK
111.682189 & 9$_{1,8}$ - 8$_{1,7}$ ($A$)  & -4.7512  &  27.9 & Unblended  \\ %OK
111.169903 & 10$_{0,10}$ - 9$_{0,9}$ ($E$)  & -4.7056  &  30.0 & Unblended$^{*}$  \\ %OK
111.171634 & 10$_{0,10}$ - 9$_{0,9}$ ($A$)  & -4.7055  &  30.0 & Unblended$^{*}$  \\ %OK
132.105508 & 12$_{1,12}$ - 11$_{1,11}$ ($E$)  & -4.4925  &  42.1 &  Unblended$^{*}$  \\ %OK
132.107205 & 12$_{1,12}$ - 11$_{1,11}$ ($A$)  & -4.4924  &  42.1 & Unblended$^{*}$  \\ %OK
132.245128 & 12$_{0,12}$ - 11$_{0,11}$ ($E$)  & -4.4915  &  42.1 & Blended: H$^{13}$CCCN \\ %OK
132.246730 & 12$_{0,12}$ - 11$_{0,11}$ ($A$)  & -4.4914  &  42.1 & Blended: H$^{13}$CCCN  \\ %OK
132.921937 & 11$_{1,10}$ - 10$_{1,9}$ ($E$)  & -4.5308  &  40.1 & Unblended  \\%OK
132.928736 & 11$_{1,10}$ - 10$_{1,9}$ ($A$)  & -4.5306  &  40.1 & Unblended  \\%OK
\hline 
\end{tabular}
\label{tab:transitionscisL1157}
{\\ (a) ``U" refers to blending with emission from an unknown (not identified) species; transitions with the $^{*}$ symbol are not blended with emission from other species, but (auto)blended with another transition of \textit{cis}-MF.
}
\end{table*}

\section{Complementary Figures}

\begin{center}
\begin{figure*}[ht]
     \centerline{\resizebox{1.0
     \hsize}{!}{\includegraphics[angle=0]{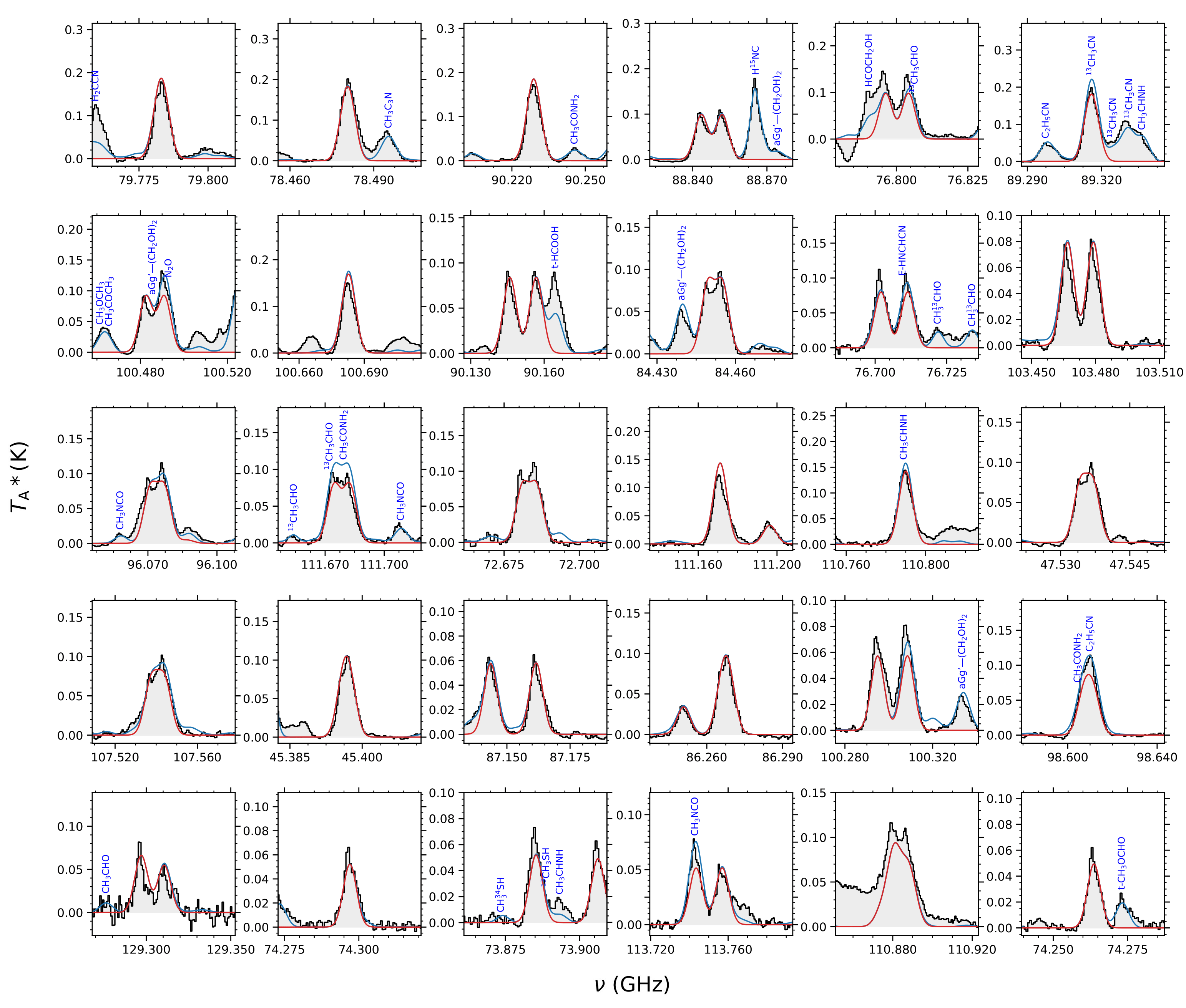}}}
     \caption{Selected transitions of \textit{cis}-MF identified toward G+0.693 (listed in Table \ref{tab:transitionscisG0693} sorted by decreasing peak intensity. The result of the best LTE fit of \textit{cis}-MF is plotted with a red line and the blue line depicts the emission from all the molecules identified to date in our survey of G+0.693 (observed spectra shown as black lines and gray-shaded histograms).}
\label{f:LTEspectrumcisG0693}
\end{figure*}
\end{center}

\begin{center}
\begin{figure*}[ht]
     \centerline{\resizebox{1.0
     \hsize}{!}{\includegraphics[angle=0]{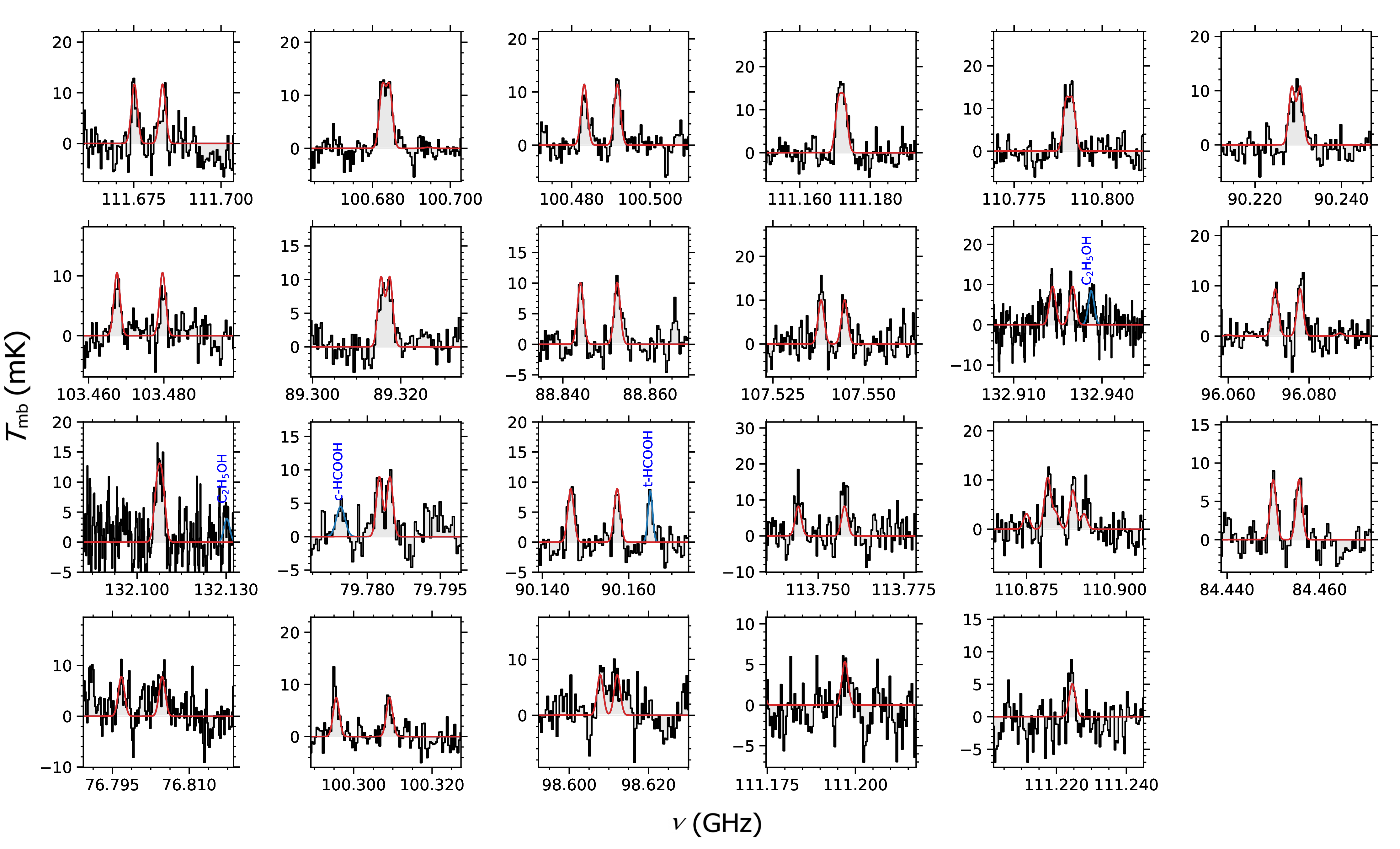}}}
     \caption{Selected transitions of \textit{cis}-MF identified toward L1157-B1 (listed in Table \ref{tab:transitionscisL1157}) sorted by decreasing peak intensity. The result of the best LTE fit of \textit{cis}-MF is plotted with a red line and the blue line depicts the emission from the molecules identified toward L1157-B1 that appear in the vicinity of the lines belonging to \textit{cis}-MF (observed spectra shown as black lines and gray-shaded histograms).}
\label{f:LTEspectrumcisL1157}
\end{figure*}
\end{center}

In Figures \ref{f:LTEspectrumcisG0693} and \ref{f:LTEspectrumcisL1157}, we present the fitted line profiles of \textit{cis}-MF toward G+0.693 and L1157-B1, respectively. We used the same color code for the line profiles of Figure \ref{f:LTEspectrumtransG0693} (i.e., fitted line profiles of \textit{cis}-MF in red and the expected molecular emission from all the molecules detected to date toward G+0.693 in blue).

\end{appendix}
\end{document}